\documentclass[aps,prl,twocolumn,groupedaddress,amsmath,amssymb]{revtex4-1}
\usepackage{graphicx}  
\usepackage{dcolumn}   
\usepackage{bm}        
\usepackage{verbatim}   

\numberwithin{equation}{section}
\begin{document}

\title{Tunneling of two bosonic atoms from a one-dimensional anharmonic trap}
\author{I.S. Ishmukhamedov$^{a,b}$ and V.S. Melezhik$^{a,c}$}
\affiliation{
   $^a$ Bogoliubov Labortory of Theoretical Physics, Joint Institute for Nuclear Research, Dubna, Moscow Region 141980, Russian Federation \\
   $^b$ Al-Farabi Kazakh National University, Almaty 050040, Kazakhstan \\
   $^c$ State University "Dubna", Dubna, Moscow Region 141980, Russian Federation
   }

\begin{abstract}
We investigate the quantum dynamics of two interacting bosonic atoms confined in a one-dimensional anharmonic trap. The tunneling rate, an experimentally measurable parameter of the system, was calculated as a function of the effective coupling interatomic constant $g$ from the ground ($n=N=0$) and first excited atomic states in the trap with respect to relative ($n=2,N=0$) and center-of-mass ($n=0,N=2$) atomic motion. This allows to investigate the initial population and pair correlation, as well as the effective coupling constant $g$, of the system by comparing the calculated tunneling rate with the experimental one. We have observed that the only possible tunneling scenario is a sequential particle tunneling in the cases we considered.
We have also analyzed a rearrangement $(0,2)\rightleftarrows(2,0)$ of the spectrum in the limit $g\rightarrow \pm 0$ of noninteracting atoms.
\end{abstract}

\maketitle

\section{I. Introduction}
One of the basic effects of quantum mechanics - a particle tunneling through a repulsive barrier, responsible for such fundamental processes as alpha decay and nuclear fission and fusion, has attracted in recent years a great attention in connection with the cold atom simulation of different phenomena from solid state to nuclear and high energy physics. Particularly, in the recent experiments\cite{zuern} such aspects of the atomic tunneling trough the walls of confining traps as pairing and BCS-BEC crossover (from a Bardeen-Cooper-Schrieffer pairing to a Bose-Einstein condensate) have been investigated. To this class of problems can also be attributed a tunneling of BEC\cite{beinke} and ultracold bosonic few-body systems\cite{lode,lode1}, a transport of the repulsive BEC and a modeling of the Josephson effect in a double-well trap\cite{nesterenko}. In the works\cite{zuern,zuern2} it was shown that the tunneling rate through the walls of atomic traps is an experimentally measurable parameter containing an important information about the atomic dynamics inside the trap as well as the initial state of the quantum system. However, to extract this information one has to perform a corresponding accurate calculation of the tunneling rate for comparison with the experimental one.

So far, a theoretical description of the tunneling dynamics through repulsive barriers of different form is quite non-trivial task. In the modern computations a semi-classical approach of Wentzel-Kramers-Brillouin (WKB) remains the basic analytical method despite the known shortcomings\cite{rontani, rontani2}. The main disadvantage here is that the WKB method completely neglects interparticle interactions and therefore can produce significant errors in the end results\cite{gharashi}. Therefore, to treat different tunneling dynamics, which depends on the specific peculiarities of each problem, a variety of numerical approaches was developed during the last two decades in atomic, molecular and nuclear physics\cite{melezhik,krass,gusev,gharashi,maruyama,scamps}.

In the present work we investigate the tunneling dynamics of two interacting bosonic atoms through the walls of a one-dimensional (1D) anharmonic trap by using an extension of the computational splitting-up technique suggested in \cite{melezhik} for ionization of hydrogen-like atoms by strong electric fields. With this approach we calculate the dependence of the tunneling rate on the effective coupling constant $g$ from the first three low-lying atomic states in a confining trap. The rates of the transitions between the states are also investigated. The obtained results can be used to recover the physical picture inside the confining trap by comparing the calculated tunneling rates with the experimental ones. Similar tunneling processes were qualitatively investigated for tunneling through a box-shaped potential model from the ground state of a rectangular potential well\cite{hunn}.

The paper is organized as follows. In Section II we define the Hamiltonian of two-atomic system confined in 1D anharmonic trap. Key points of the splitting-up method are given in Section III for numerical integration of the 2D time-dependent Schr\"{o}dinger equation describing two-body quantum dynamics in a 1D anharmonic atomic trap. Special attention is paid here for the stable and accurate procedure of extracting the tunneling rate of the system. The obtained results of calculation of the tunneling rates and transition probabilities are given and discussed in Section IV. Here, we also discuss the rearrangement of the spectrum of the confined two atoms in the limit $g\rightarrow \pm 0$ of noninteracting atoms. Finally, in Section V we draw our conclusions and provide a short outlook.

\section{II. Problem formulation}
The quantum dynamics of two identical bosonic atoms with masses $m$ in 1D confining trap $\sum\limits_{j=1,2}V(x_j)$ is described by the following Hamiltonian:
\begin{eqnarray}
\nonumber
H=&-&\frac{\hbar^2}{2m}\frac{\partial^2}{\partial x_1^2}-\frac{\hbar^2}{2m}\frac{\partial^2}{\partial x_2^2}+V(x_1)+V(x_2)
\\
&+&V_{\textrm{int}}(x_1-x_2)\,\,,
\end{eqnarray}
where the interatomic potential $V_{\textrm{int}}(x_1-x_2)$ is chosen in the Gaussian form
\begin{eqnarray}
V_{\textrm{int}}(x_1-x_2)=-V_0\exp\left\{-\frac{(x_1-x_2)^2}{2r_0^2}\right\}
\end{eqnarray}
with the depth $V_0$ and $r_0$ defining the range of the interaction.

The Hamiltonian (2.1) can be considered as an effective Hamiltonian describing dynamics of two atoms tightly confined in the transverse direction $(y,z)$ (atomic motion is forbidden in the transverse direction) but with a softer confinement in the longitudinal direction $x$ described by a standing-wave form \cite{peng,bloch}:
\begin{eqnarray}
V(x_j)=V_\textmd{d}\sin^2\left(\frac{2\pi}{\lambda}x_j\right),~~~j=1,2.
\end{eqnarray}
The interaction of the atom $j$ with the optical trap (2.3) is defined by the wavelength $\lambda$ of the external laser field and the atomic polarizability (included in the parameter $V_\textmd{d}$) \cite{peng,bloch}. Here we use the parametrization of the works\cite{peng,ishmukh}
\begin{eqnarray}
V(x_j)=-\frac{\hbar\omega}{12\alpha}\sin^2\left(\sqrt{-6\alpha}\dfrac{x_j}{\ell} \right),~~~j=1,2\,\,,
\end{eqnarray}
where the parameter of anhamonicity $\alpha=-\dfrac{8\pi^2 \hbar}{12\lambda^2 m \omega}$, and $\omega$ and $\ell$, defined as $\omega=\dfrac{2\pi}{\lambda}\sqrt{\frac{2|V_\textmd{d}|}{m}}$ and $\ell=\sqrt{\frac{\hbar}{m\omega}}$, were introduced.

To have a realistic scale for the atom-trap interaction (2.4) we use the parameters $\lambda$ and $\omega$ corresponding to the optical traps from the experiment\cite{haller}, where the confined $^{133}$Cs atoms were investigated (see Table 1).
\begin{center}
\begin{tabular}{cc}
 \hline \hline
Trap frequency, $\omega$ & $2\pi \times 14.5$ kHz \\
Wavelength, $\lambda$ & $1.06449 \times 10^{-4}$ cm \\
Anharmonicity, $\alpha$ & $-0.0304552$ \\ \hline \hline
\end{tabular}\\
Table 1. The trap parameters from the Innsbruck experiment\cite{haller}.
\end{center}

In the present work we restrict ourselves by the consideration of the atomic dynamics in the single-well of the 1D lattice (2.4) by approximating the latter as
\begin{eqnarray}
\nonumber
V^{(\textsf{sw})}(x_j)
=&&
\begin{cases}
-\frac{\hbar\omega}{12\alpha}\sin^2\left(\sqrt{-6\alpha}\dfrac{x_j}{\ell} \right), \hspace{.3cm} |x_j|\leq\frac{\pi \ell}{\sqrt{-6\alpha}} \\
0, \hspace{3.55cm} |x_j|>\frac{\pi \ell}{\sqrt{-6\alpha}}
\end{cases}
\\
&&j=1,2
\end{eqnarray}
Such approximation neglects the tunneling of the atoms through the neighbour walls as well as the reflection from the walls (see Fig.1)
\begin{center}
\includegraphics[scale=0.25]{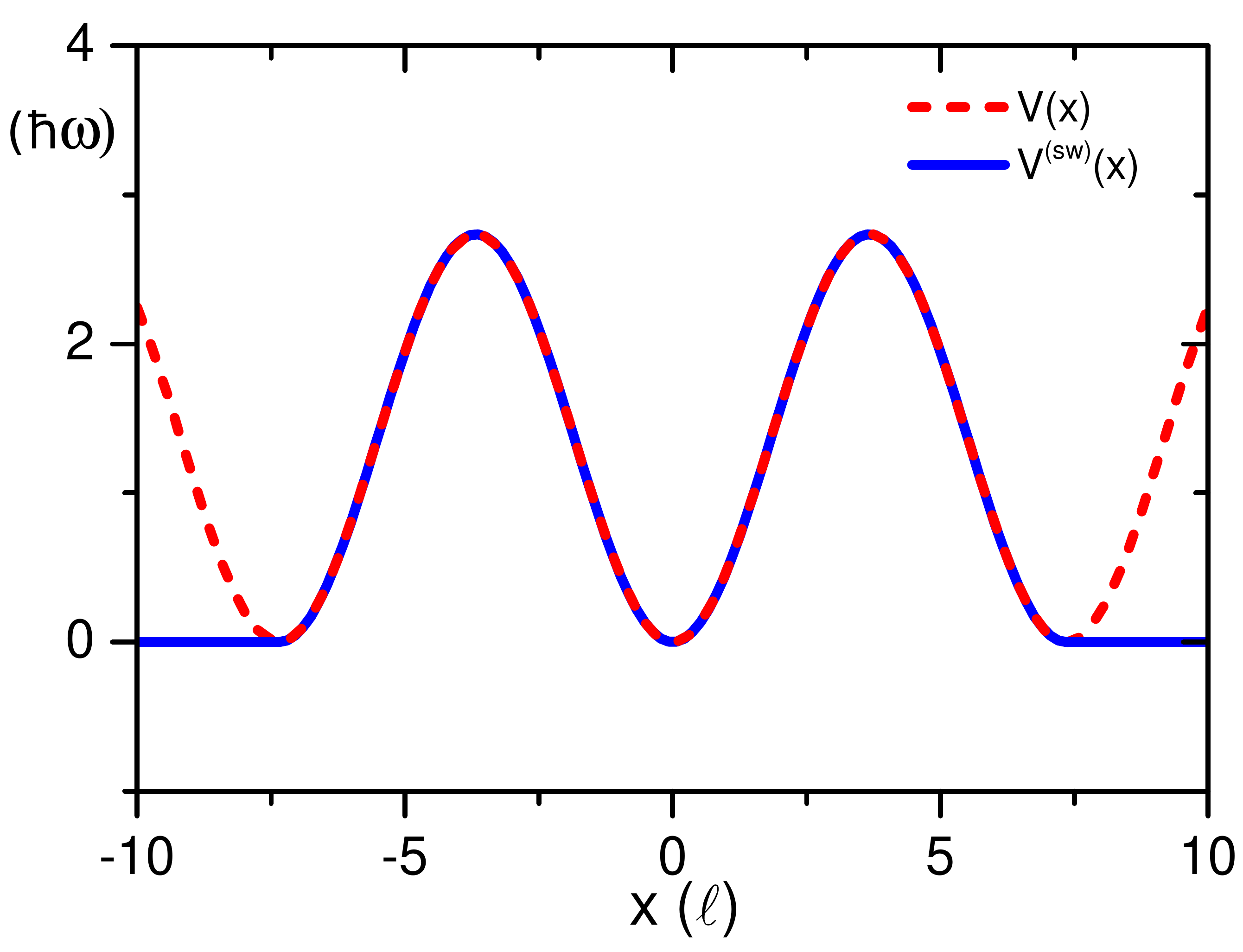}\\
Fig.1: (Color online) Approximation (2.5) (gray solid line) of the exact trap potential (2.4) (blue dashed line) for $\alpha=-0.0304552$.
\end{center}

\section{III. Method}
\setcounter{section}{3}
\setcounter{equation}{0}
To calculate a tunneling rate, $\gamma$, from the bound state of the potential $V_{\textrm{int}}(x_1-x_2)+ \sum\limits_{j=1,2}V^{(\textsf{sw})}(x_j)$ we integrate the 2D time-dependent Schr\"{o}dinger equation (SE)
\begin{eqnarray}
i\hbar\frac{\partial \psi(x_1,x_2,t)}{\partial t}=H(x_1,x_2)\psi(x_1,x_2,t)
\end{eqnarray}
with the Hamiltonian $H(x_1,x_2)$ defined by Eqs.(2.1), (2.2) and (2.5).
Based on ideas from \cite{marchuk}, which were developed in the works\cite{melezhik2,melezhik3} in application to confined ultracold atom-atom collisions in waveguide-like traps,  we employ the component-by-component split-operator method to integrate Eq.(3.1):
\begin{eqnarray}
\nonumber
&&\psi(x_1,x_2,t+\Delta t)=\exp\left\{-i \frac{\Delta t}{2\hbar} V_{\textrm{int}}(x_1-x_2)\right\}
\\ \nonumber
&&\times\exp\left\{-\frac{i \Delta t H_1(x_1)}{\hbar}\right\}
\exp\left\{-\frac{i \Delta t H_2(x_2)}{\hbar}\right\}
\\
&&\times
\exp\left\{-i \frac{\Delta t}{2\hbar} V_{\textrm{int}}(x_1-x_2)\right\}
\\ \nonumber
&&\times \psi(x_1,x_2,t)
\end{eqnarray}
where
\begin{eqnarray}
H_j(x_j)=-\frac{\hbar^2}{2m}\frac{\partial^2}{\partial x_j^2}+V^{(\textsf{sw})}(x_j),\hspace{.5cm}j=1,2.
\end{eqnarray}
The computational scheme (3.2) is correct up to terms of the $\mathcal{O}(\Delta t^3)$ order. Following \cite{melezhik2,melezhik3}, we approximate the action of the differential operators $\exp\left\{-i \Delta t/\hbar H_j(x_j)\right\}$ by implicit Crank-Nicolson scheme
{\small
\begin{eqnarray}
\nonumber
\exp\left\{-\frac{i \Delta t H_j(x_j)}{\hbar}\right\}=\left(1+\frac{1}{2\hbar}i H_j\Delta t\right)^{-1}&&\left(1-\frac{1}{2\hbar}i H_j\Delta t\right)\,,
\end{eqnarray}
}
which maintains the accuracy of the split-operator method (3.2).

Finiteness of the width of the confining potential wall causes broadening of the energy levels in the potential describing interatomic and atom-trap interactions $V_{\textrm{int}}(x_1-x_2)+\sum\limits_{j=1,2}V^{(\textsf{sw})}(x_j)$ due to the atom tunneling through the wall of the confining trap. This means that we need to have an outgoing wave away from the region of action of the confining potential, i.e. at $x_1,x_2\rightarrow \pm \infty$ \cite{meyer, maruyama}. This kind of boundary condition in the time-dependent scheme can be modeled by introducing at the edge of the radial grid $x_m$ some type of an \textit{absorber} \cite{meyer} or a mask function \cite{kulander}. Here we choose the scheme with the additional complex absorbing potential (CAP) near the edge of the radial grid in the form suggested in \cite{meyer, scamps}
\begin{eqnarray}
W(x_j)=w_c(|x_j|-x_c)^2\theta(|x_j|-x_c),\hspace{.2cm}j=1,2,
\end{eqnarray}
where $\theta(x)$ - is the Heaviside step function and the parameter $x_c$ defining the region where the CAP switches on and it should be chosen at the point behind the barrier of the confining potential $V^{(\textsf{sw})}$. The choice of the parameter $w_c$ is discussed in the paragraph after Eq.(3.6).

A plot of CAP (3.4) with the confining potential $V^{(\textsf{sw})}$ (2.5) is shown in Fig.2.
\begin{center}
\includegraphics[scale=0.25]{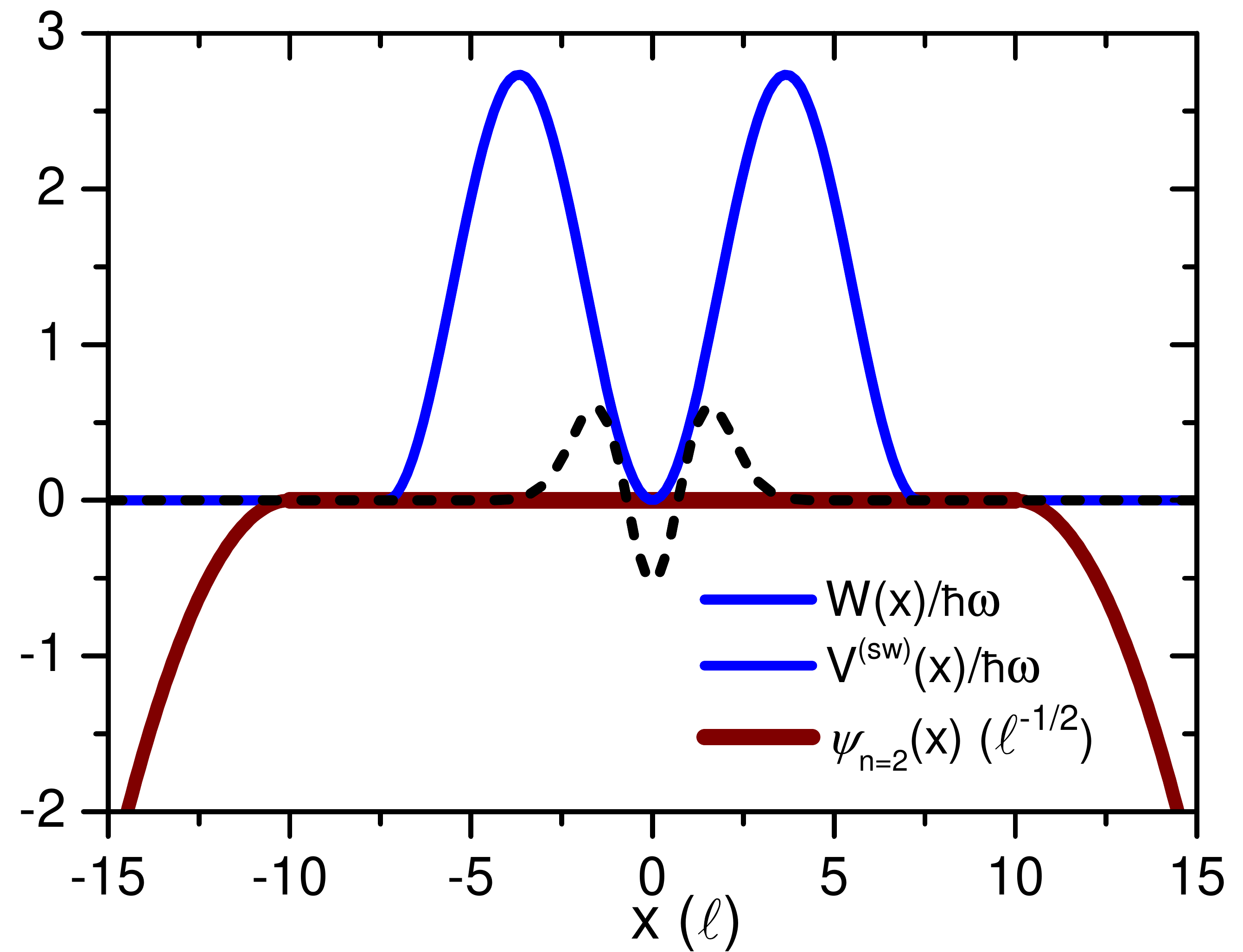}\\
Fig.2: (Color online) Plot of the absorbing potential $W(x)$ (3.4) (thick red solid line) and the confining potential $V^{(\textsf{sw})}(x)$ (2.5) for $\alpha=-0.0304552$ (blue solid line). We also plot the wave function, $\psi_{n=2}(x)$ (black dashed line) of the first excited state in the confining potential for illustrating the scale of the problem.
\end{center}

The numerical integration of the Schr\"{o}dinger equation (3.1) with the Hamiltonian $H(x_1,x_2)=\sum\limits_{j=1,2}(H_j(x_j)+iW(x_j))+V_{\textrm{int}}(x_1-x_2)$ defined by (3.3), (2.5), (3.4) and (2.2) permits to extract the desired tunneling rate  $\gamma$ (or the energy level width $\Gamma=\hbar\gamma$) from the decay of the total probability
\begin{eqnarray}\hspace{-1cm}
P(t)=\int\limits_{-x_m}^{x_m}\int\limits_{-x_m}^{x_m}
dx_1dx_2|\psi(x_1,x_2,t)|^2\sim \exp\left\{-\gamma t\right\},
\end{eqnarray}
to find the atoms in the box $|x_1,x_2|\leq x_m$, i.e. the total population of the atomic bound states in the box.
From Eq.(3.5) one can define the tunneling rate as
\begin{eqnarray}
\gamma=-\frac{1}{P(t)}\frac{dP(t)}{dt}.
\end{eqnarray}

The determination of the tunneling rate from (3.6) is obviously holds only for the exponential decay of the  probability (3.5). This condition restricts the choice of the parameters $x_c$ and $w_c$ of CAP (3.4) as well as the time domain where the decay of the norm (3.5) is stabilized after the beginning of the tunneling \cite{melezhik}. Here we choose $x_c=10\ell$ and $w_c=-0.1\hbar\omega\ell^{-2}$ so that the tunneling rate, $\gamma$, remains constant to a good accuracy.

Following the pioneering works\cite{olshanii,busch}, laid the foundation for investigations of the confined two-body systems in quasi-1D geometry of atomic traps, we define here the interatomic interaction through the effective coupling constant $g$ connected with the 1D scattering length $a_{\textmd{1D}}$ as
\begin{eqnarray}
g=-\frac{2\hbar^2}{ma_{\textmd{1D}}}.
\end{eqnarray}
The scattering length $a_{\textmd{1D}}$ was calculated by the integration of the 1D Schr\"{o}dinger equation
\begin{eqnarray}
\hspace{-1cm}
\left[
-\frac{\hbar^2}{2\mu}\frac{d^2}{dx^2}-V_0\exp\left\{-\frac{x^2}{2r_0^2}\right\}
\right]
\psi_{\textrm{sc}}(x)
=\frac{\hbar^2 k^2}{2\mu}\psi_{\textrm{sc}}(x),
\end{eqnarray}
describing the atom-atom collision in 1D free-space, with the boundary condition
\begin{eqnarray}
\psi_{\textrm{sc}}(x)\xrightarrow[x \to \pm\infty]{}\cos(k|x|+\delta(k))
\end{eqnarray}
at zero energy limit $k=\sqrt{2\mu E}/\hbar\rightarrow 0$. The calculated scattering phase $\delta(k)$ defines at $k\rightarrow 0$ the 1D scattering length
\begin{eqnarray}
a_{\textmd{1D}}=\lim_{k\rightarrow0}\frac{\cot(\delta(k))}{k}\,.
\end{eqnarray}
Here, $\mu=m/2$  and $x=x_1-x_2$ are the reduced mass and the relative coordinate of the atomic pair, respectively.

The dependence of the coupling constant $g$ on the depth $V_0$ of the interaction potential (2.2), calculated at $r_0=0.1\ell$ is shown in Fig.3. The choice of the parameter $r_0$, fixed in \cite{ishmukh2,gharashi}, adequately corresponds to current experiments \cite{zuern,zuern2,haller} where the range of confining potential $\ell$ always essentially exceeds the range of interatomic interaction: $r_0 \ll \ell$.
\begin{center}
\includegraphics[width=.45\textwidth]{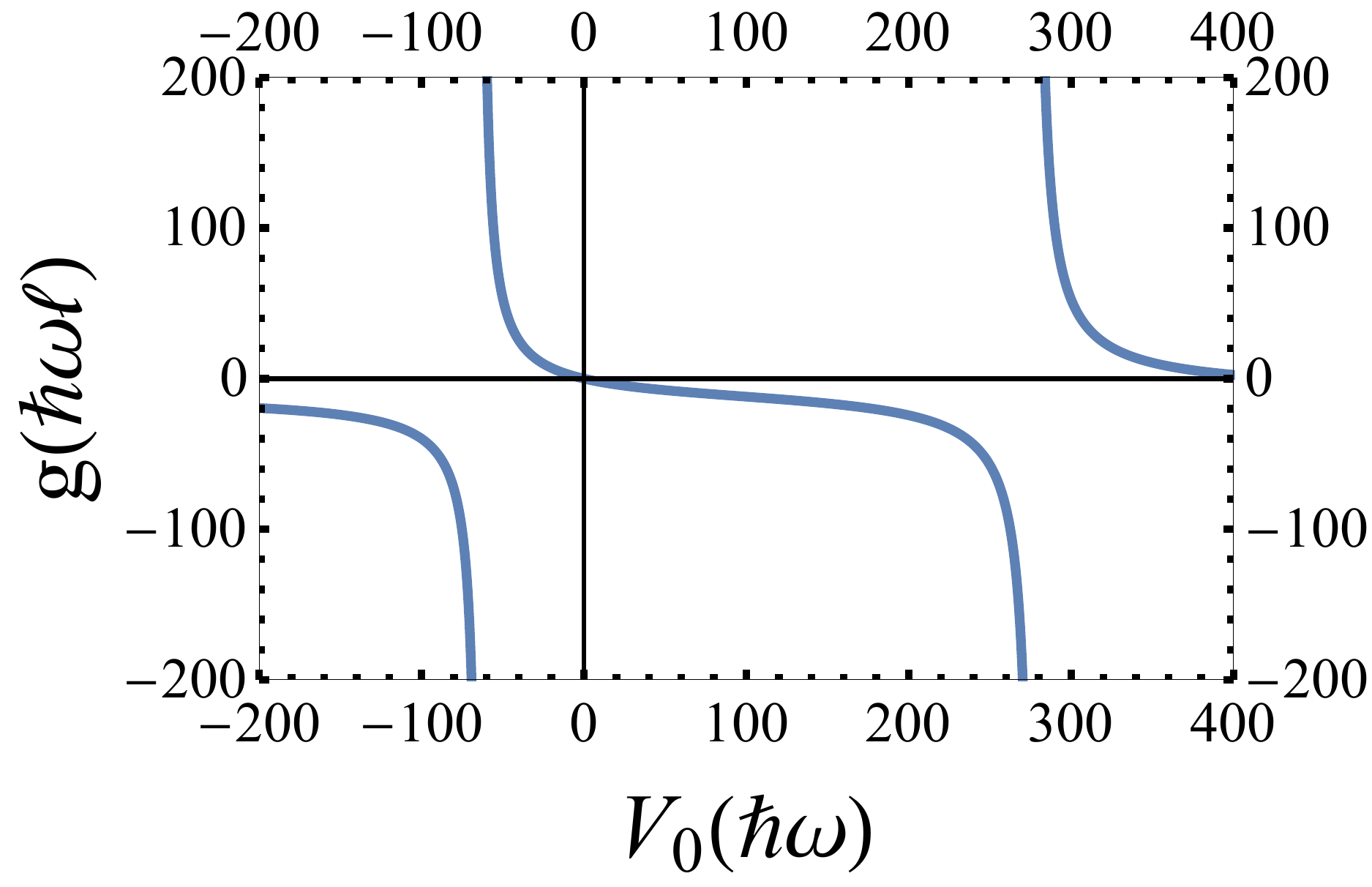}\\
Fig.3: (Color online) Dependence of the effective coupling constant $g$ on the depth $V_\textmd{0}$ of the interaction potential (2.2) (blue solid line) at $r_0=0.1\ell$.
\end{center}
In the calculations the parameter $V_0$  varied in the range $-64 \lesssim V_0/(\hbar\omega) \lesssim 53$ which corresponds to $-8 \lesssim g/(\hbar\omega\ell) \lesssim +\infty$. It has permitted to investigate the tunneling processes in a wide range of the coupling strength, $g$, from an attraction $g < 0$ to a strong repulsion $g\rightarrow +\infty$.

\section{IV. Results}
\setcounter{section}{4}
\setcounter{equation}{0}
\subsection{4.1. Preparation of the initial state}
The modern experimental set up permits preparation of the well defined and practically non-decaying initial atomic states in confining traps with the subsequent ``switch on'' of the tunneling process by means of narrowing the width of the confining potential\cite{zuern,zuern2}. To model such process\cite{gharashi}, first, we prepare the non-decaying initial atomic bound state at $t\leq0$ by solving the eigenvalue problem for the potential $V_{\textrm{int}}(x_1-x_2)+\sum\limits_{j=1,2}V^{(\textsf{6})}(x_j)$ with the confining trap\cite{grish,sala}
\begin{eqnarray}
\nonumber
V^{(6)}(x_j)&=&\hbar \omega \left(
\frac{1}{2}\left(\frac{x_j}{\ell}\right)^2+\alpha  \left(\frac{x_j}{\ell}\right)^4
+\frac{4 \alpha ^2}{5}\left(\frac{x_j}{\ell}\right)^6 \right),
\\&&\hspace{4cm}j=1,2
\end{eqnarray}
having infinite width of the walls and repeating the form of the internal part of the confining potential (2.5) $V^{(\textsf{sw})}(x_j)$. At $t>0$, the trap $V^{(6)}(x_j)$ is replaced by $V^{(sw)}(x_j)$ to allow the atoms to tunnel out of it.

A plot of the potentials $V^{(\textsf{sw})}(x_j)$ and $V^{(6)}(x_j)$ is presented in Fig.4.
\begin{center}
\includegraphics[scale=0.25]{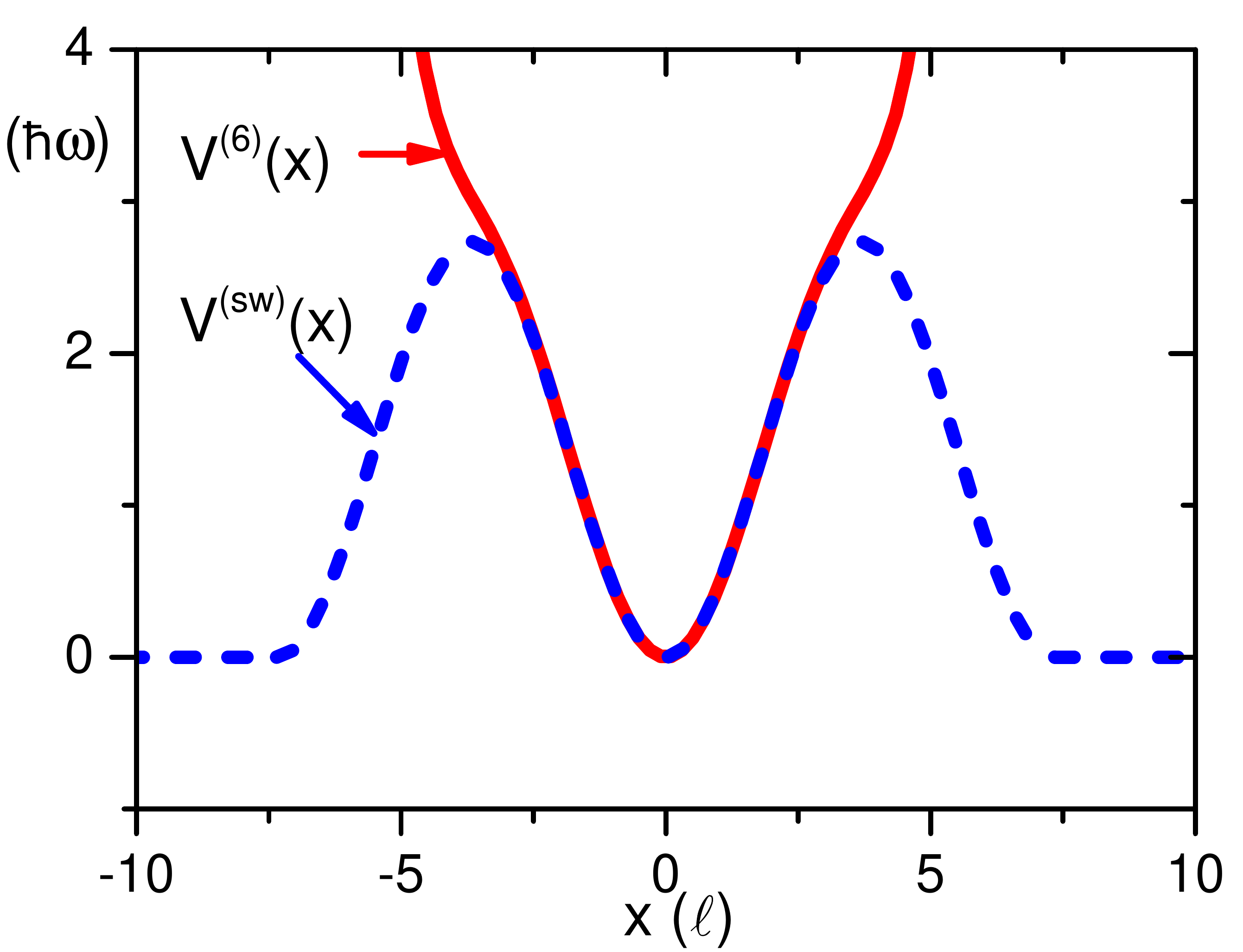}\\
Fig.4: (Color online) ``Initial'' confining trap $V^{(6)}(x)$ (red solid curve), that confines the motions of the atoms at $t\leq0$ and a modified trap $V^{(\textsf{sw})}(x)$ (blue dashed curve), which releases the atoms due to the quantum tunneling at $t>0$.
\end{center}

\begin{center}
\includegraphics[scale=0.3]{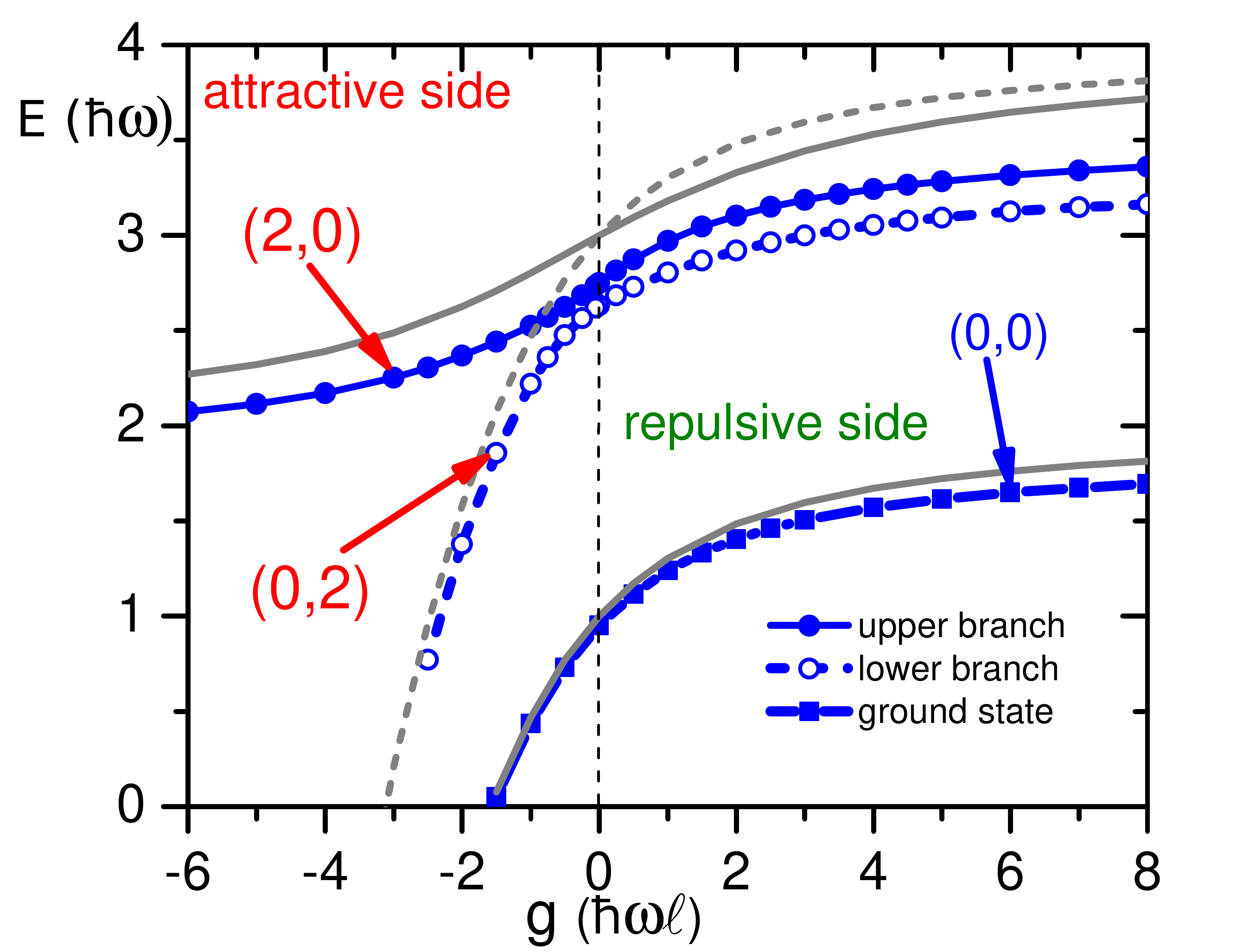}\\
Fig.5: (Color online) Three lowest energy levels of two bosonic atoms as a function of the coupling constant $g$ in the confining potential $V^{(6)}(x_j)$ with $\alpha=-0.0304552$ (blue curves) and in the harmonic trap $V^{(2)}(x_j)=\frac{1}{2}m\omega^2x_j^2$ (gray curves).
\end{center}

In Fig.5 we present the energy spectrum of two bosonic atoms calculated as a function of the coupling constant $g$ in the confining potential $V^{(6)}(x_j)$ (4.1) and in the harmonic trap $V^{(2)}(x_j)$ corresponding the case $\alpha=0$.

For identification of the calculated states of the spectrum we use quantum numbers $n$ and $N$ characterizing the quantization (induced by the trap) of the relative atomic and the center-of-mass motions, respectively, which are good ones in the harmonic limit (i.e. when $\alpha=0$ in (4.1)) due to the separation of the relative atomic and center-of-mass variables.

In Fig. 5 the three lowest states of the calculated spectrum are presented: the ground state ($n=0,N=0$) and the first two excited states ($0,2$) and ($2,0$).

When $\alpha=0$ the energy branches $(0,2)$ and $(2,0)$ cross each other at zero coupling, $g=0$ (non-interacting atoms). This corresponds to a pure two-dimensional harmonic oscillator and these levels become degenerate due to rotational symmetry. This symmetry breaks if $\alpha\neq0$ and we observe an avoided crossing of these energy levels at $g=0$.

These states rearrange when they cross the point $g=0$, that is, the state $(2,0)$ turns into the state $(0,2)$ and, vice versa, the state $(2,0)$ turns into the $(0,2)$ state (see Section 4.3). The quantum numbers (n,N) of the two-atomic state inside of the trap we define by the nodal structure of the initial wave function $\psi_{n,N}(x,y,t=0)$ with respect to variables of the relative motion $x=x_1-x_2$ and the center-of-mass $y=(x_1+x_2)/2$. The analysis in Sections 4.3 shows that the first excited state at negative $g$ is (0,2) and the second one is (2,0) (see Fig.5). At positive $g$, the first excited state becomes (2,0) and the second one - (0,2) due to the rearrangement $(0,2)\rightleftarrows(2,0)$ of the spectrum in the limit $g\rightarrow \pm 0$.

\subsection{4.2. Tunneling dynamics from bound states of the two-atomic confined system}

By numerical integrating the 2D time-dependent Schr\"{o}dinger equation (3.1) for $t>0$ we calculate the time evolution of the two-atomic wave-packet (3.2) from the ground ($0,0$) and excited states ($0,2$) and ($2,0$) as a function of the coupling constant $g$.
In the calculation the sixth-order finite-difference approximation on the uniform spacial grid over $x_1$ and $x_2$ was used\cite{melezhik2,melezhik3}. The range, $|x_j|\leq x_m$, of the space of radial variables was chosen as $x_m=20\ell$ and the step of integration over radial variables as well as the step of integration over the time $\Delta t=0.01\omega^{-1}$ were chosen to keep the accuracy of the calculation of the tunneling rates within the order of one percent.
In Fig.6 we present the calculated tunneling rates $\gamma$ defined by (3.6) from the first two excited states ($0,2$) and ($2,0$) for a wide range of the strength of the coupling constant $g$.
\onecolumngrid
\begin{center}
\includegraphics[width=.6\textwidth]{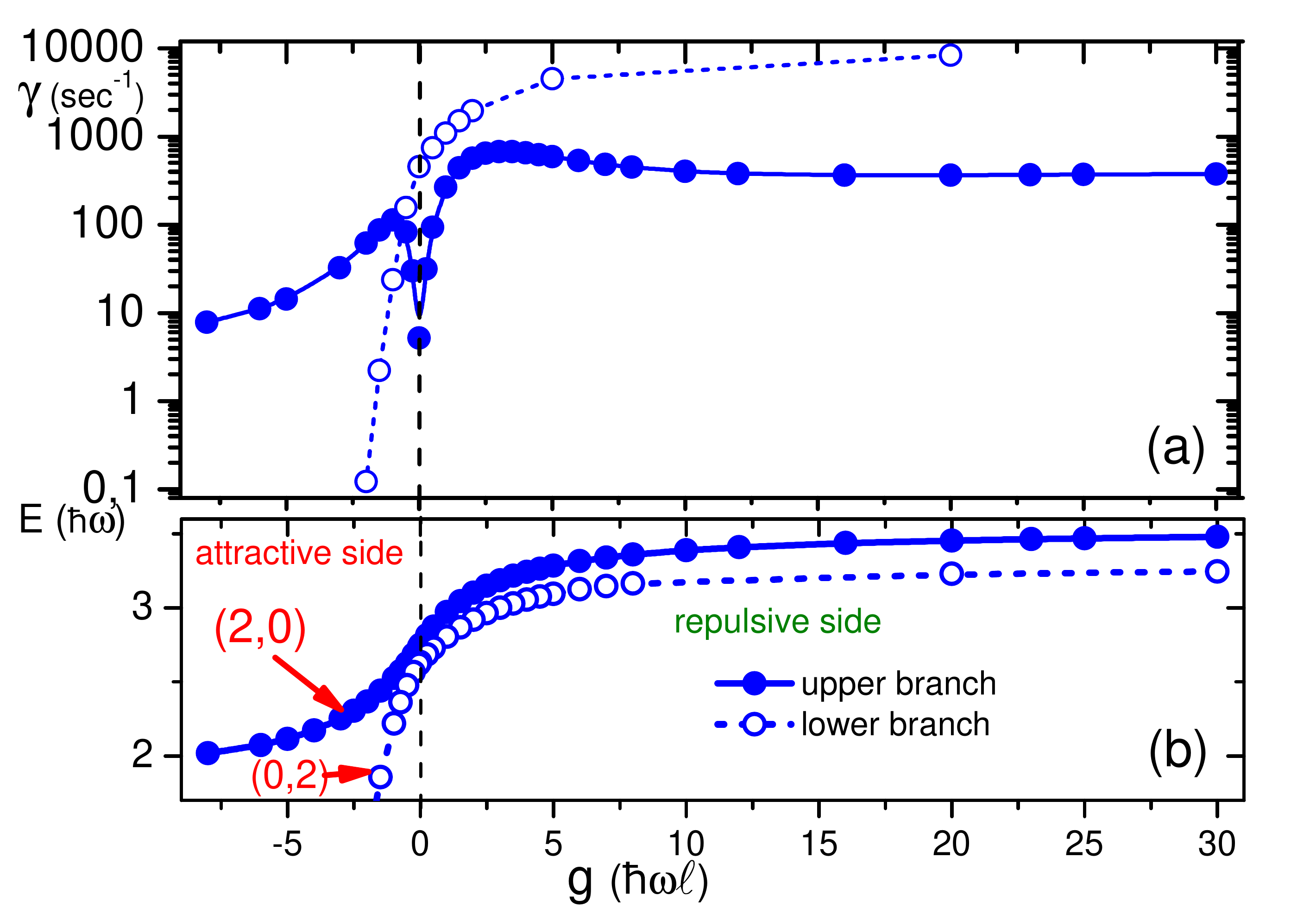}\\
Fig.6: (Color online) (a) Tunneling rates from the first two excited states as a function of the coupling constant $g$ for $\alpha=-0.0304552$. (b) The binding energies of the first two excited states as a function of the coupling constant $g$ for $\alpha=-0.0304552$.
\end{center}
\twocolumngrid
\begin{center}
\textit{Upper energy branch of the excited states}
\end{center}

Fig.7 demonstrates the calculated time-evolution of the total probabilities $P(t)$ (3.5) decaying from the upper branch of energy levels for excited states presented in Fig.6,b and the corresponding tunneling rates $\gamma(t)$ (3.6). This figure demonstrates quite fast transition of the decay of the total probabilities $P(t)$ to the exponential law as well as fast convergence of the tunneling rates $\gamma(t)$ to the limiting value $\gamma(t) \rightarrow \gamma(\infty)$ for a wide range of the coupling constant $g$.

To understand a mechanism of tunneling from the upper energy branch of the excited states we also calculate the probability flux $j_k(x_1,x_2,t)$
\begin{eqnarray}
\nonumber
j_k(x_1,x_2,t)=\frac{\hbar}{2mi}\left( \psi^{\ast}\frac{\partial \psi}{\partial x_k}-\psi\frac{\partial \psi^{\ast}}{\partial x_k} \right),
\\
k=1,2
\end{eqnarray}
at certain time, $t$. In Fig.8 the fluxes $|\textbf{j}(x_1,x_2,t)|$, calculated for different values of the coupling constant $g$, are presented.
\begin{center}
\includegraphics[scale=0.3]{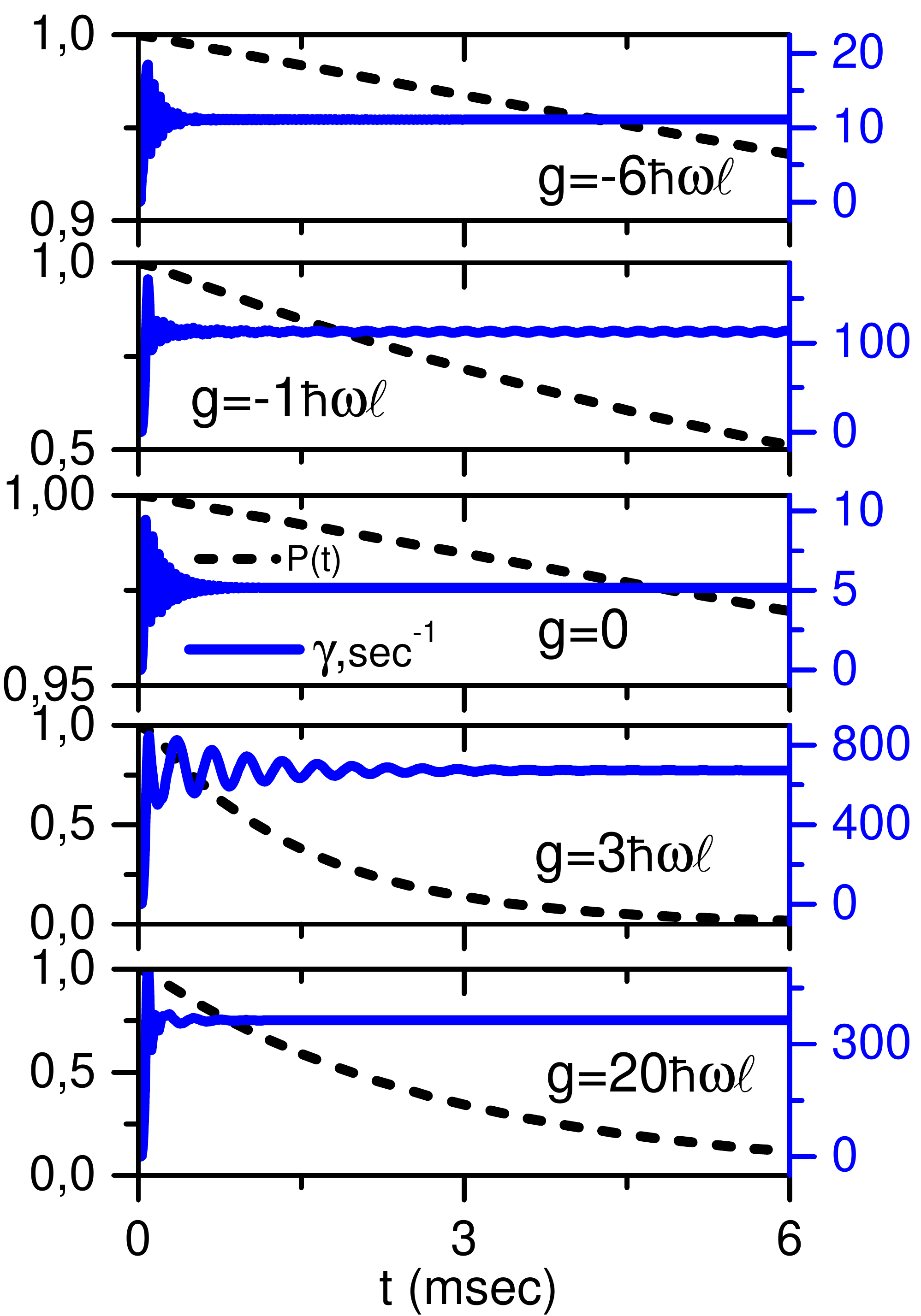}\\
Fig.7: (Color online) The time-evolution of the total probability $P(t)$ (black dashed curves) from the upper branch of energy levels for excited states presented in Fig.6,b  and the corresponding tunneling rates $\gamma$ (blue solid curves). Calculations were performed for a few values of the coupling constant: $g/(\hbar\omega \ell)=-6,-1,0,3,20$.
\end{center}
\onecolumngrid
\begin{center}
\begin{tabular}{ccc}
\hspace{-1.2cm}
$g=-6\hbar\omega\ell$ &\hspace{-1.2cm} $g=0$ &\hspace{-1.2cm} $g=20\hbar\omega\ell$ \\
\includegraphics[width=.32\textwidth]{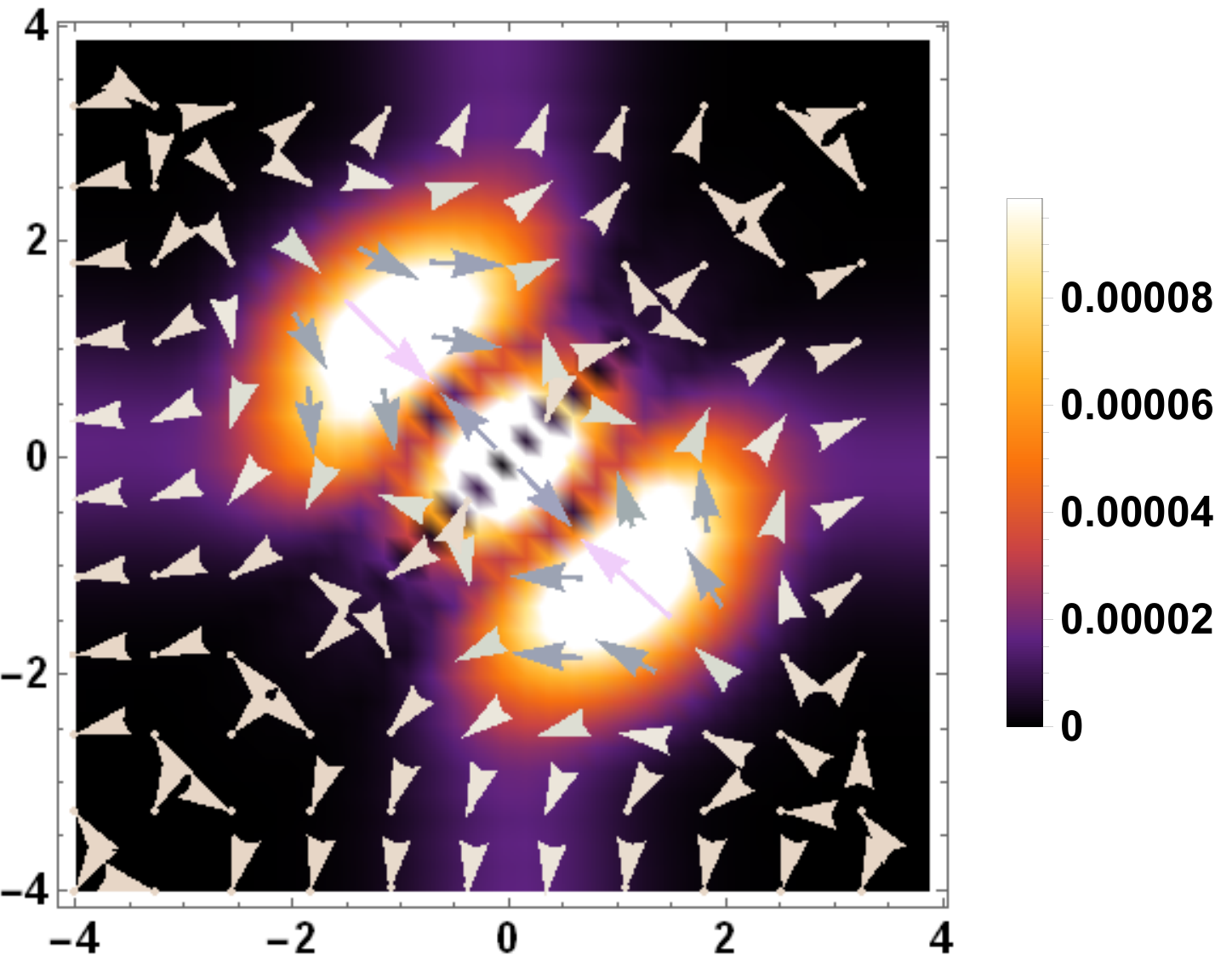} &
\includegraphics[width=.32\textwidth]{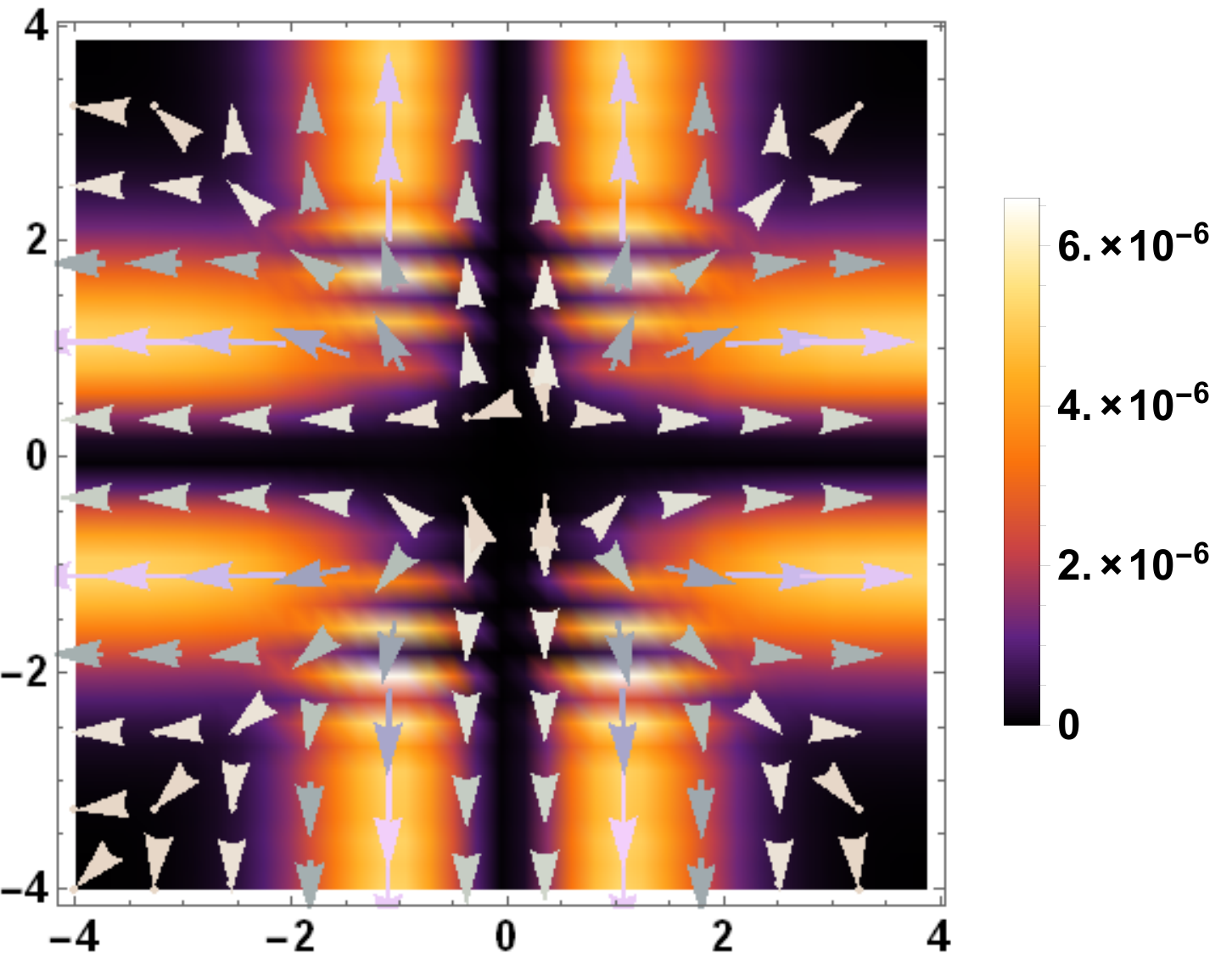} &
\includegraphics[width=.32\textwidth]{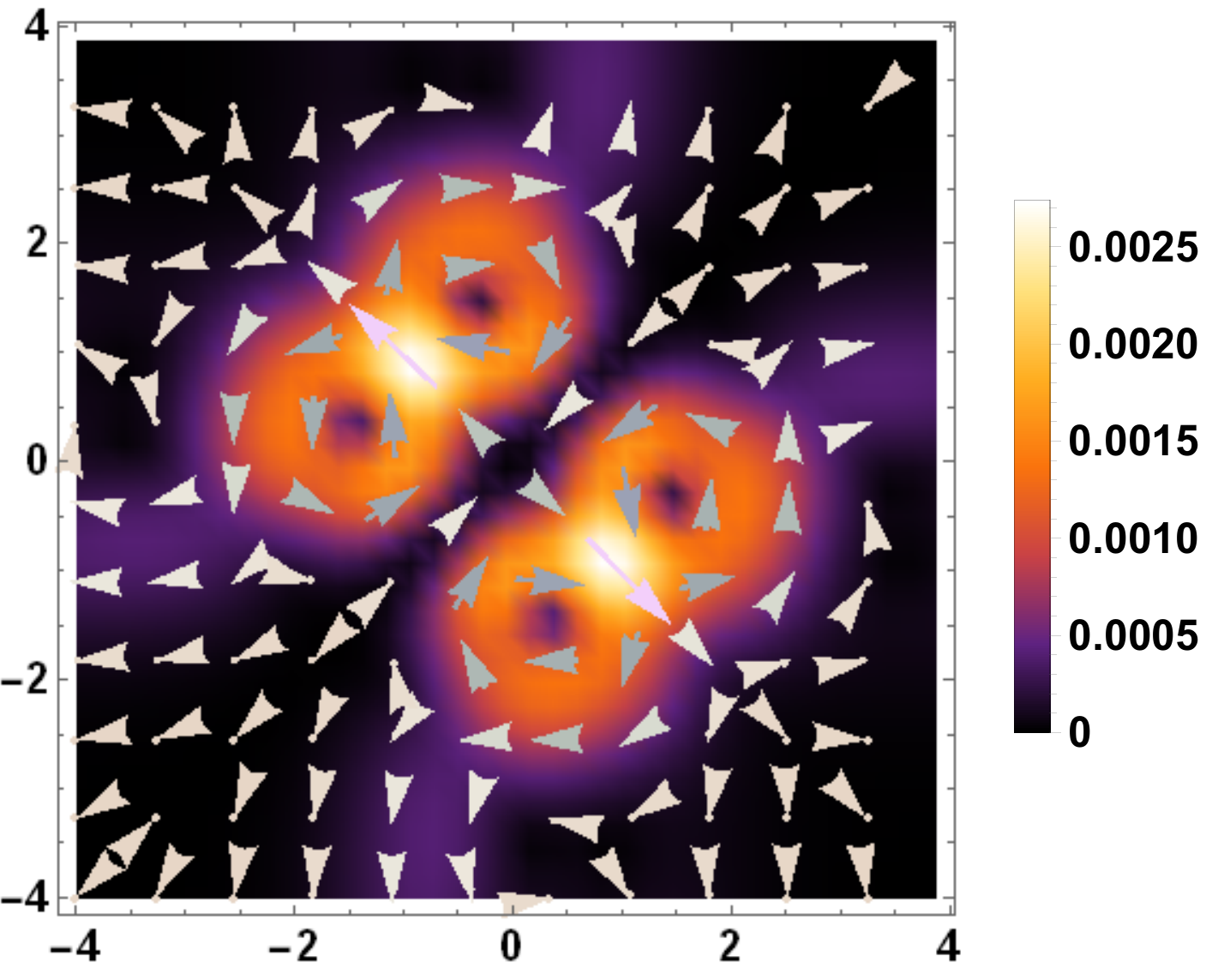} \\
\end{tabular}\\
Fig.8. Distribution of a modulus of the flux $|\textbf{j}(x_1,x_2,t)|$ (in $\omega \ell^{-1}$ units) from the upper branch of the excited states presented in Fig.6 and its direction (in arbitrary units) for different values of the coupling constant $g$ at $t=120\omega^{-1}\approx1.32$msec.
\end{center}
\twocolumngrid

From Fig.8 we see that the flux in all of the cases shows a complicated behavior near the origin $x_1=x_2=0$. What all graphs in Fig.8 have in common is that in all considered cases the dominate flux directions are the directions along the axes $x_1$ and $x_2$ what corresponds to a single-particle tunneling.

To analyze the mechanism of the tunneling more quantitatively we divide (following the work \cite{maruyama}) the whole space of radial variables ($x_1,x_2$) into several regions (see Fig.9).
\begin{center}
\includegraphics[scale=0.3]{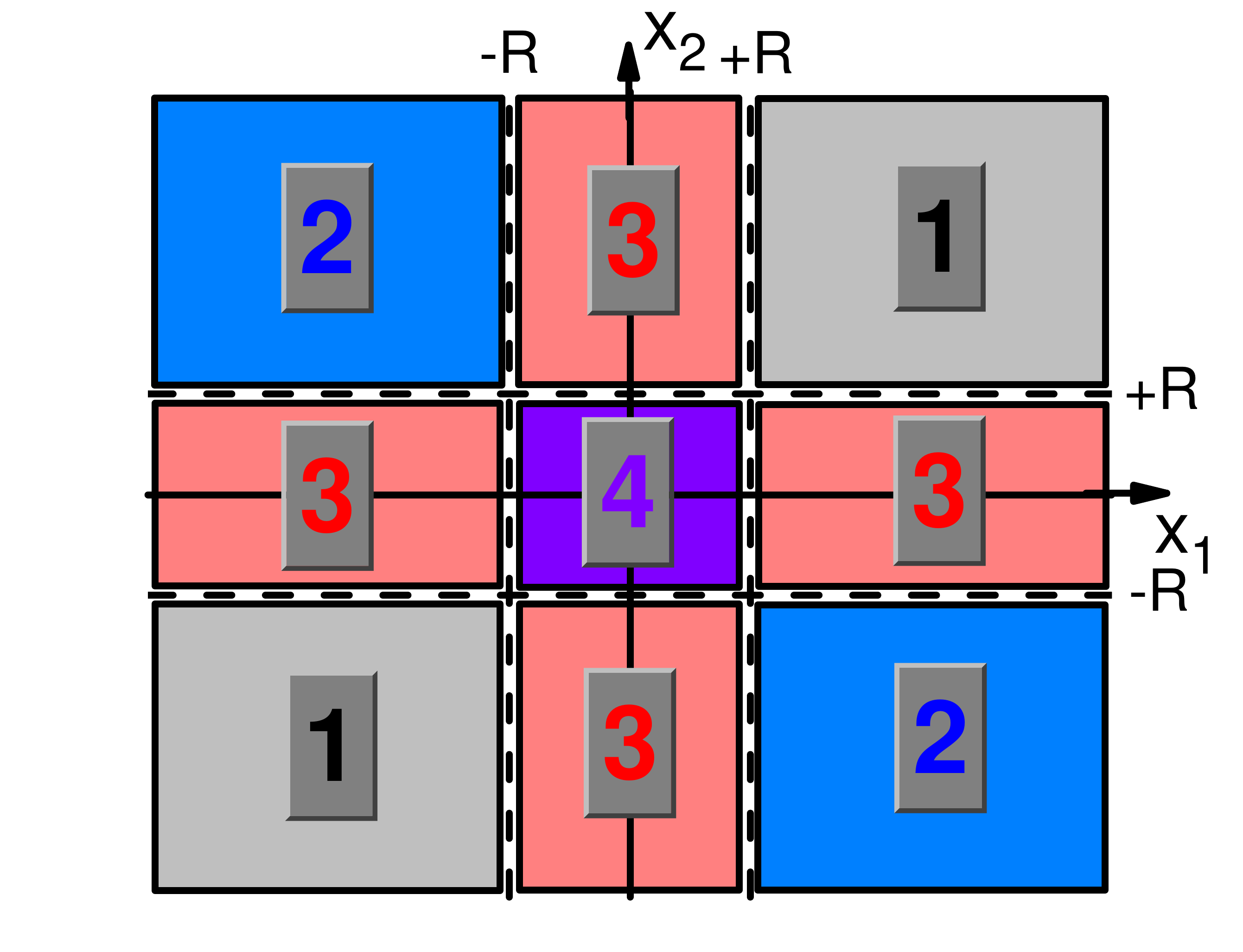}\\
Fig.9: Division of the radial variable space into several regions for the partial probabilities $P_k$ analysis.
\end{center}
and calculate partial probabilities, $P_k$, to find the atom or atomic pair in each region $k$
\begin{eqnarray}
\nonumber
P_k(t)=\iint\limits_{region~k} dx_1 dx_2|\Psi(x_1,x_2,t)|^2\,,
\\
k=1,2,3,4\,.
\end{eqnarray}

Detection of the atom in the regions $3$ corresponds to a single-particle tunneling from the trap and in the regions $1$ corresponds to a tunneling of two particles as a bound system. The situation in which a particle tunnels to the regions $2$ was observed in \cite{maruyama}. We define the size of the region $4$ to approximately cover the initial atomic distribution and take $R=5\ell$.

The calculated time evolution of the partial probabilities $P_k(t)$ is shown in Fig.10.
\begin{center}
\includegraphics[width=.35\textwidth]{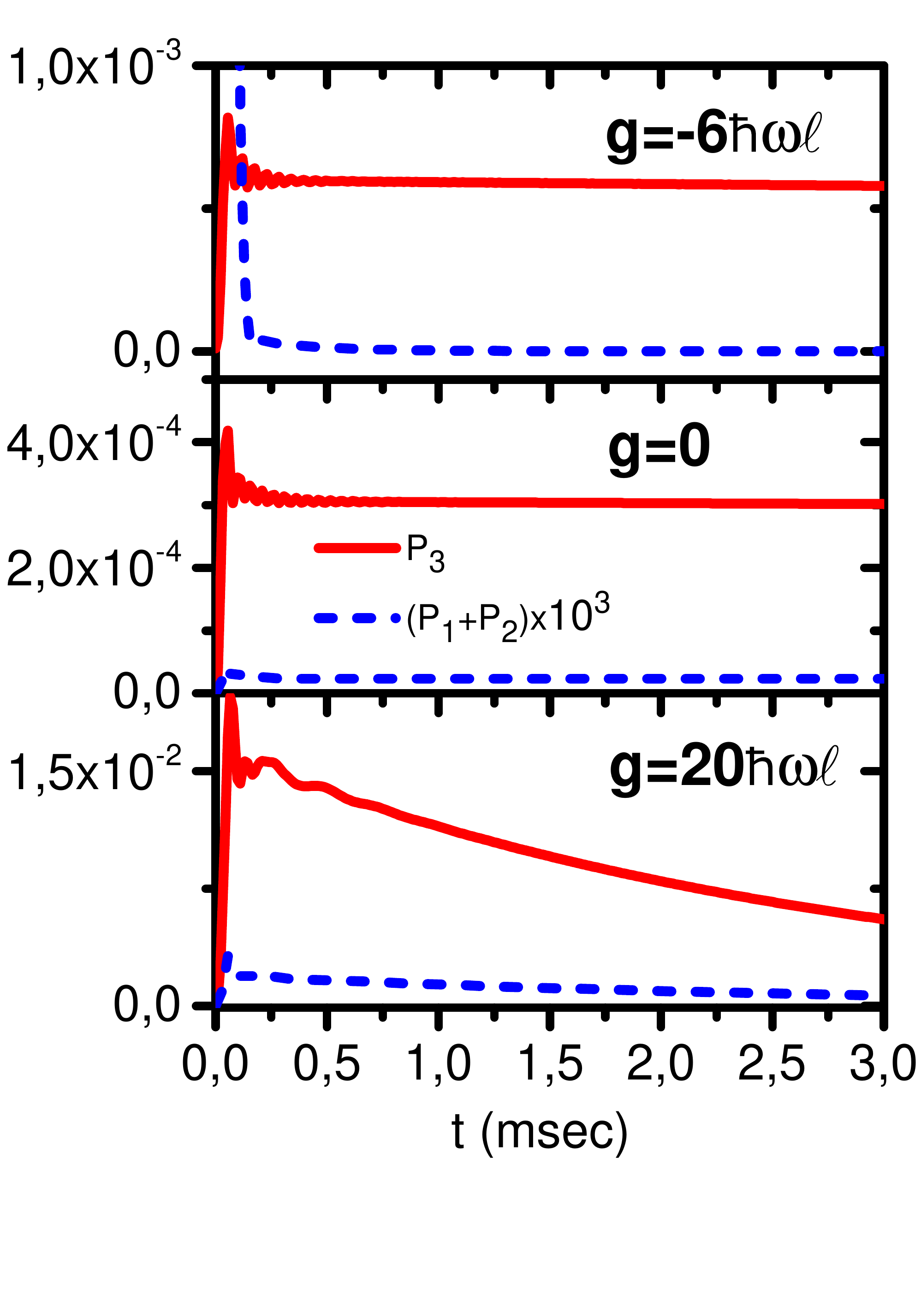}\\
Fig.10: (Color online) Partial probabilities $P_k(t)$ (4.3) to populate the regions $k$ for three different values of the coupling constant $g$.
\end{center}
Fig.10 demonstrates that the partial probabilities - $P_3(t)$ - are a few order of magnitude larger than $P_1(t)+P_2(t)$, which are practically negligible in all of the considered cases. This clearly shows that the sequential particle tunneling is the dominating mechanism of the tunneling.

\begin{center}
\textit{Lower energy branch of the excited states}
\end{center}
To understand the dynamics of a tunneling process of the atoms bound initially in the lower energy branch of the excited states (Fig.6,b) we analyze in details the case $g/(\hbar\omega\ell)=5$ since it captures all the features of such tunneling (see Fig.11).
\begin{center}
\includegraphics[scale=0.25]{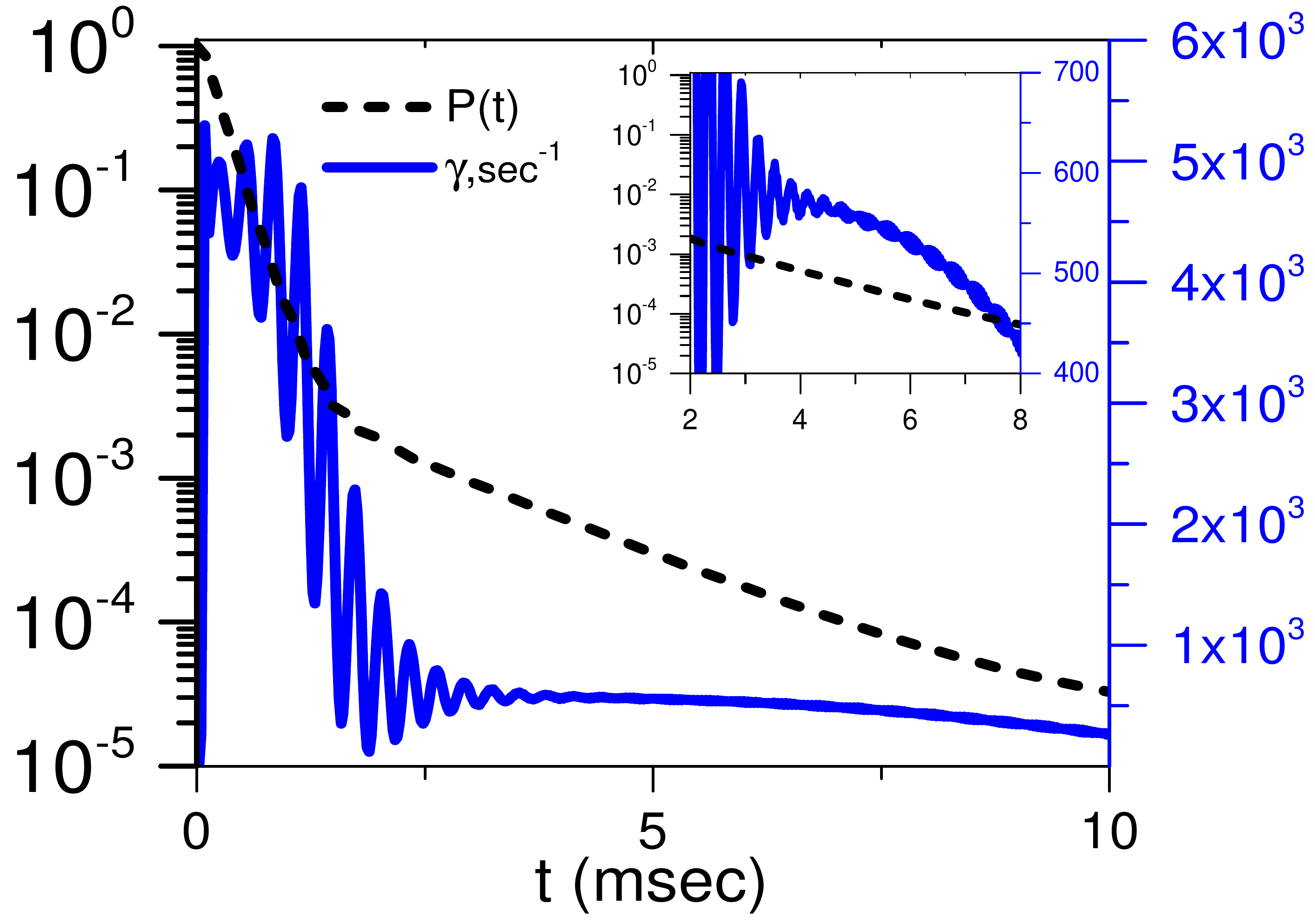}\\
Fig.11: (Color online) The tunneling rate $\gamma$ (blue solid curve) from the lower energy branch of the excited states (shown in Fig.6,b) and the corresponding total probability $P(t)$ (black dashed curve) at $g/(\hbar\omega\ell)=5$ and $\alpha=-0.0304552$. The inset shows a close-up view of the $\gamma$ and $P(t)$.
\end{center}
From Fig.11 we clearly see that the total probability $P(t)$ decreases in two stages, i.e. we see that the initial exponential behavior of $P(t)$ decreases as $\exp\left\{ -\gamma_1 t\right\}$ which turns to, approximately at $1.5-2.5$ msec, the same exponential law, but now with different tunneling rate $\gamma_1 \rightarrow \gamma_2$. The tunneling rate $\gamma_1$ calculated with Eq.(3.7) highly oscillates during the first stage of the decay and after approximately $\simeq1.5-2.5$ msec, when the second stage of the decay becomes dominating, the oscillations in $\gamma_2(t)$ significantly damp out.

To extract the tunneling rate $\gamma_1$ we fit the total probability $P(t)$ to the exponential function at the time-interval $t\leq 1.5-2.5$ msec (Fig.12)
\begin{eqnarray}
P_\textmd{fit}=P_0e^{-\gamma_1t}
\end{eqnarray}
\begin{center}
\includegraphics[scale=0.25]{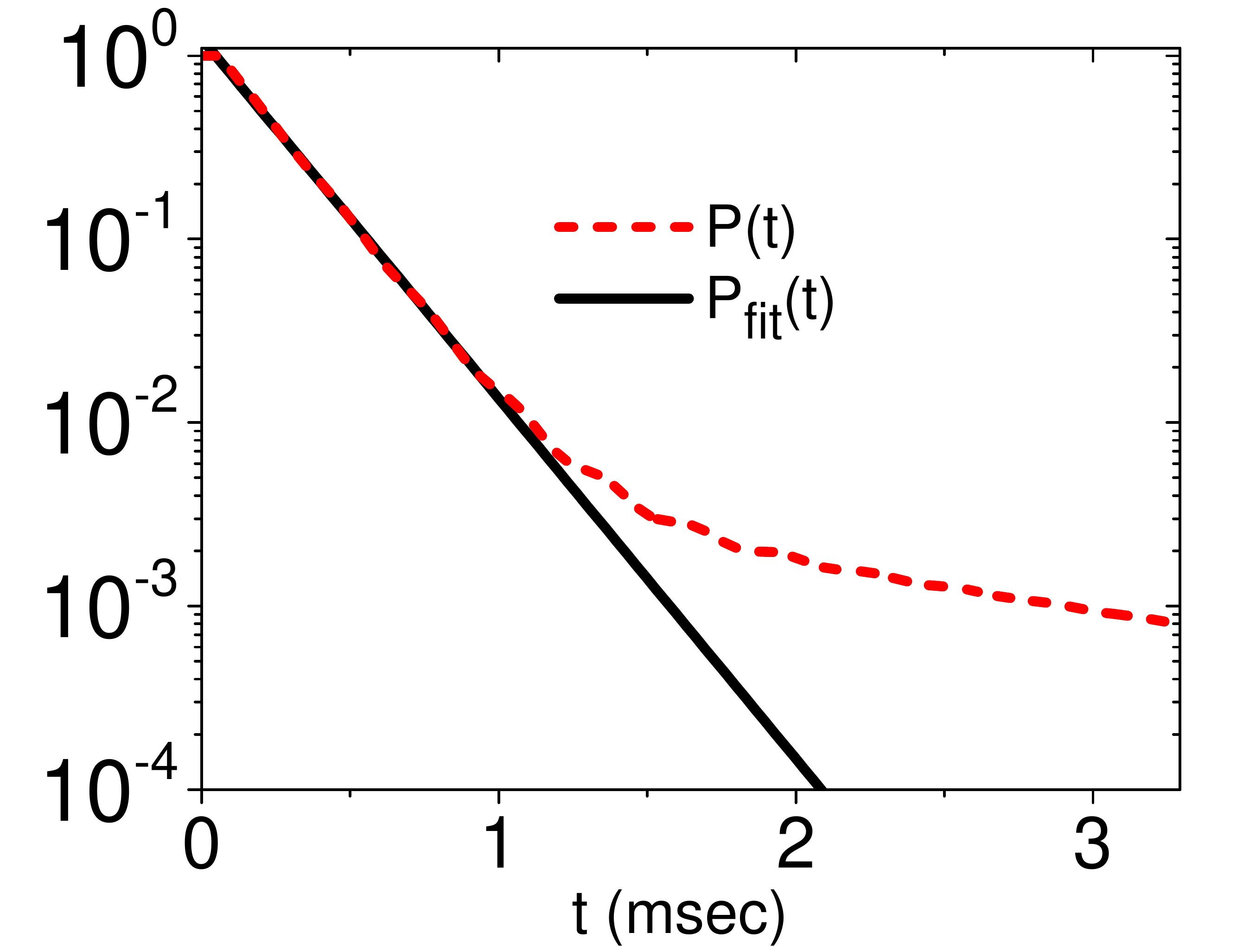}\\
Fig.12: (Color online) The total probability $P(t)$ (black dashed curve) and a fitting function (4.4) (red dashed line) in a logarithmic scale. The tunneling rate extracted from the fitting function is about $\gamma_1\simeq4520$sec$^{-1}$ in the region $t\leq1$ msec.
\end{center}
By using this fitting procedure we extract the tunneling rate $\gamma_1$ of the first decay stage for a wide range of coupling constant $g$. The result of calculation of the tunneling rate $\gamma_1$ is plotted in Fig.6,a with open circles.

The tunneling rate, $\gamma_2$, of the second stage of the decay from the lower energy branch of the excited states converges to a constant value with growing the time much better than $\gamma_1$ but not as good as the tunneling rates from the upper energy branch given in Fig.7, which can be noticed from the inset of Fig.11. If we fit the total probability $P(t)$ with the exponential function in a window $2-10$ msec we get for the $\gamma_2\simeq580$ sec$^{-1}$. This value approximately coincides with the value of tunneling rate from the upper energy branch which equals $\simeq592$ sec$^{-1}$ at $g/(\hbar\omega\ell)=5$ (see tunneling rates in Fig.6,a labeled with closed circles). To understand this effect we have calculated the time-evolution of the populations of the first three low-lying bound states of the two-atomic confined system (see Fig.13) by formulaes
\begin{eqnarray}
\nonumber
P_L(t)=|\langle \psi(x_1,x_2,t)|\psi^{(L)}(x_1,x_2) \rangle|^2 \\ \nonumber
P_U(t)=|\langle \psi(x_1,x_2,t)|\psi^{(U)}(x_1,x_2) \rangle|^2 \\
P_G(t)=|\langle \psi(x_1,x_2,t)|\psi^{(G)}(x_1,x_2) \rangle|^2\,,
\end{eqnarray}
where  $\psi^{(L)},\psi^{(U)}$, and $\psi^{(G)}$ are the wave functions of the two-atomic bound states corresponding to the lower and upper energy branches of the excited states and the ground-state in confined geometry of the trap (4.1).
\begin{center}
\includegraphics[scale=0.25]{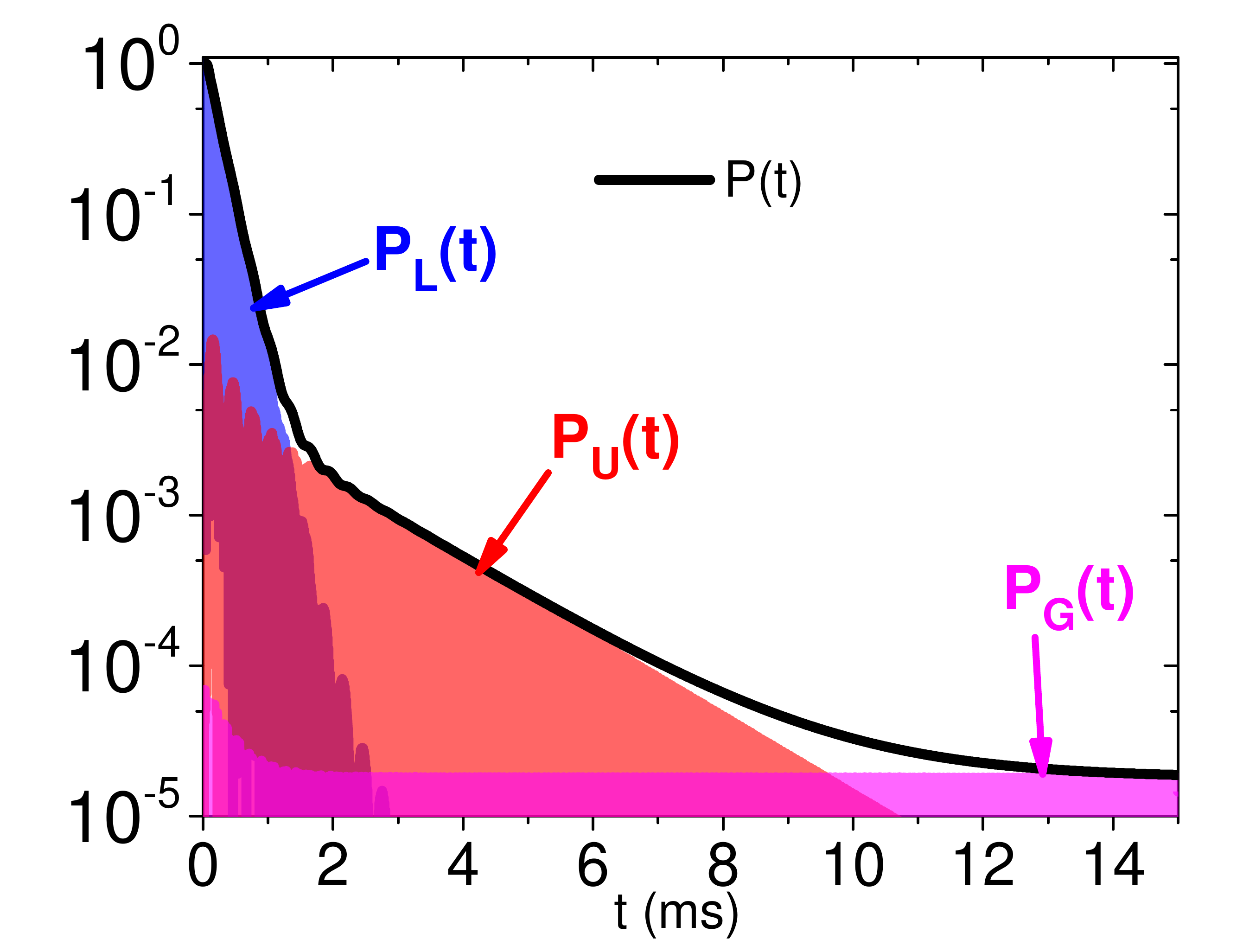}\\
Fig.13: (Color online) Time-evolution of the populations  $P_L(t),P_U(t)$, and $P_G(t)$ and the total probability $P(t)$ (black solid curve) to find atoms in the trap calculated at $g/(\hbar\omega\ell)$.\\
\end{center}
From Fig.13 one can notice that after approximately $1.5$ msec, the population $P_U(t)$ of the upper energy branch becomes dominating due to transition from the lower energy branch. That is, after about $1.5$ msec the tunneling occurs from the upper energy branch and therefore the value $\gamma_2$ of the tunneling rate approaches to the tunneling rate from the upper energy branch calculated in the previous subsection.

With increasing time to about $10$ msec the populations of the upper energy branch and the ground states become comparable (see Fig.13) and with further increasing of time the system passes to the ground state where the tunneling rate naturally defined by the decay of the ground state. It is interesting that in the case of very strong interatomic coupling $g=\infty$ the second stage goes with the tunneling rate which approximately coincides with the tunneling rate from the initial ground state (Fig.14) due to more fast population of the ground state than the upper energy branch. To see that, one has to compare the inset of Fig.14 with the converged result for the tunneling rate $\gamma$ from the ground state given in Fig.16.
\begin{center}
\includegraphics[scale=0.3]{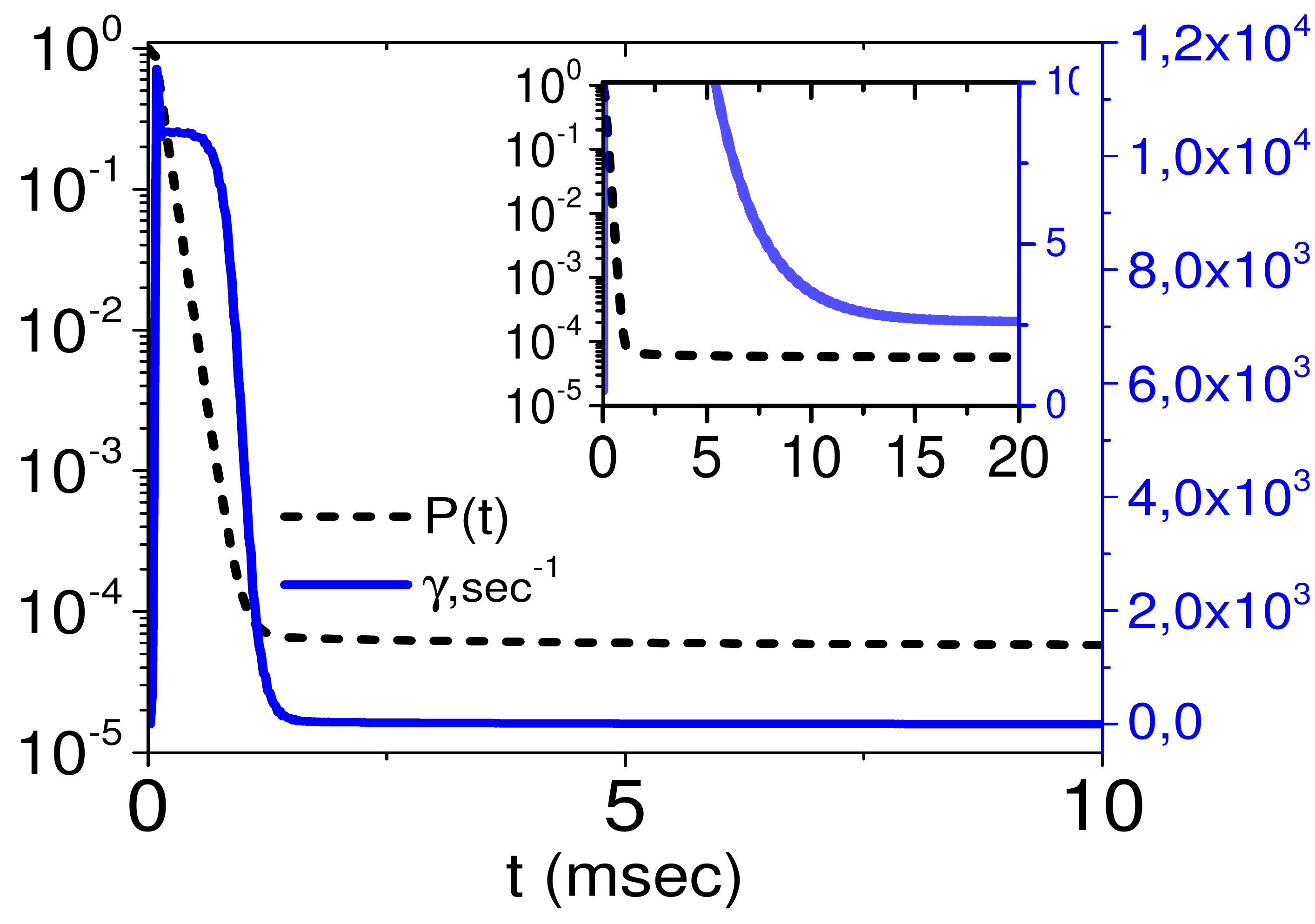}\\
Fig.14. (Color online) The tunneling rate, $\gamma$, (blue solid curve) from lower branch of the excited state (shown in Fig.6,b) at $g\rightarrow\infty$ and the corresponding total probability $P(t)$ (black dashed curve). The inset shows the convergence $\gamma(t)\rightarrow\gamma(\infty)$ to the value which approximately coincides with $\gamma$ from the ground state.
\end{center}

\begin{center}
\textit{Ground state}
\end{center}

The tunneling rate from the ground state behaves monotonically and it is significantly smaller in magnitude than those from the excited states considered above (Fig.15.a). Moreover, the tunneling is significantly suppressed even at infinite coupling constant $g=\infty$ (Fig.16). This is understood due to considerable increase of the width of the confining potential for the ground state in comparison with excited states.
\begin{center}
\includegraphics[scale=0.3]{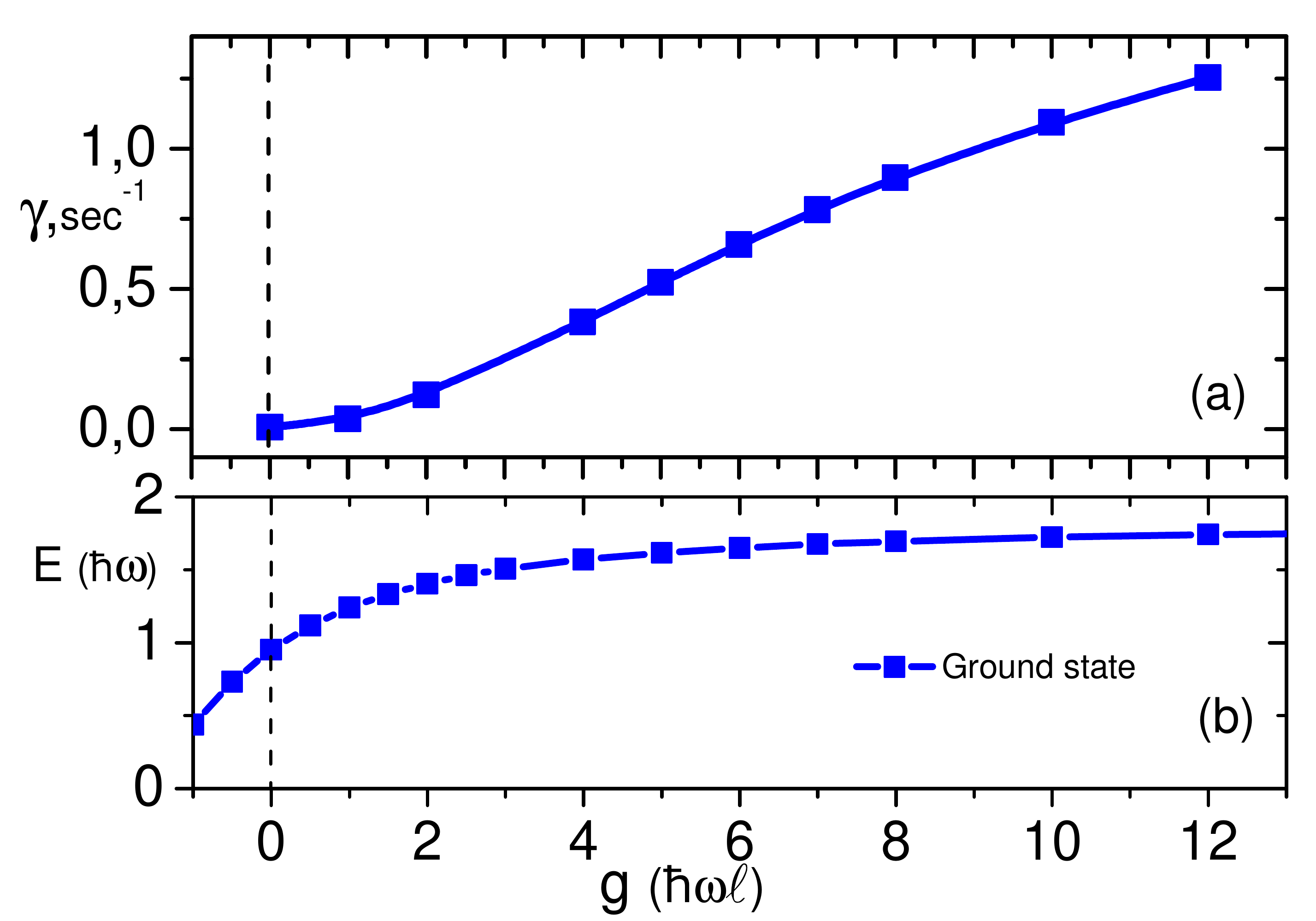}\\
Fig.15: (a) The tunneling rate $\gamma$ from the ground state as a function of the coupling constant $g$ and (b) the corresponding binding energy.
\end{center}
\begin{center}
\includegraphics[scale=0.25]{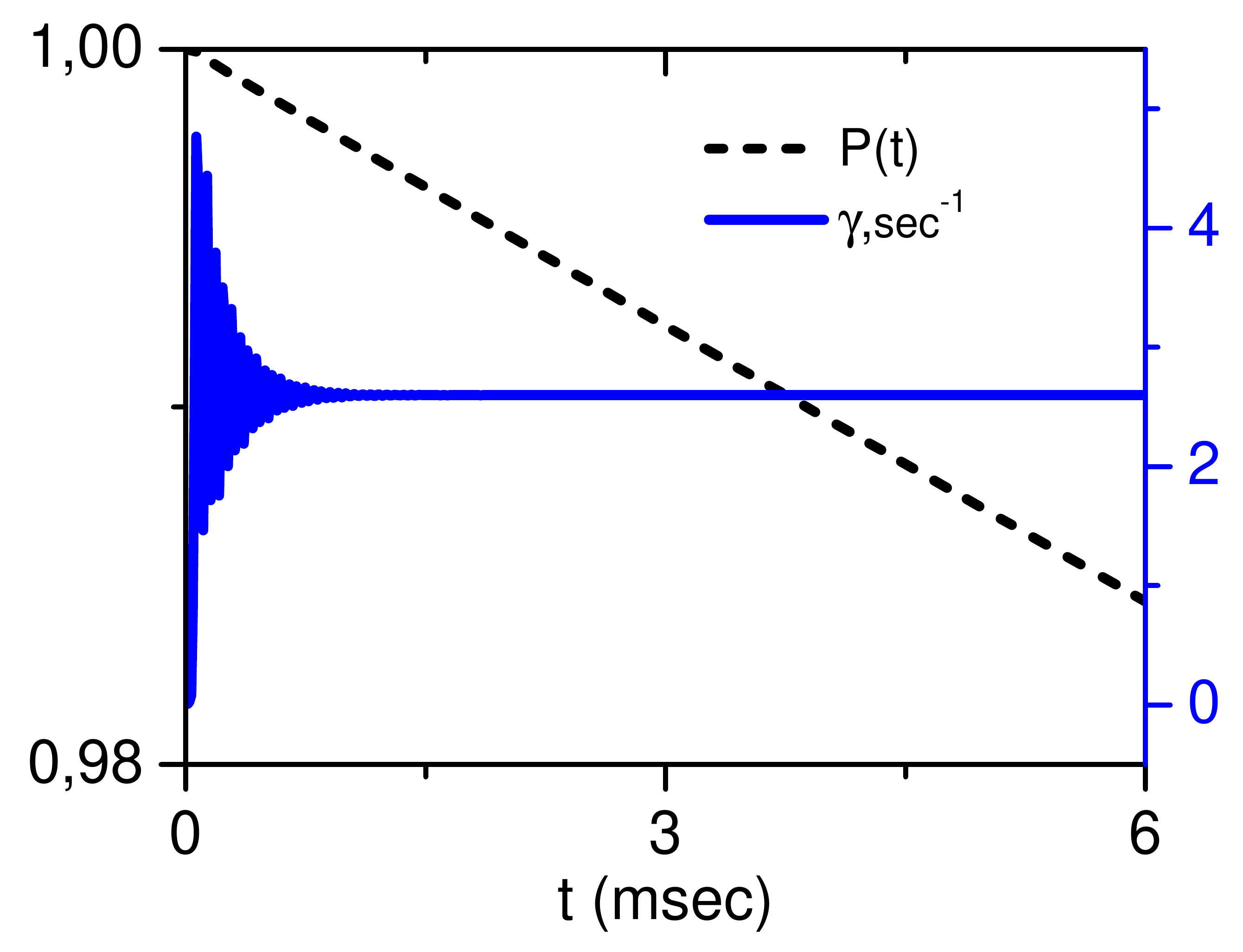}\\
Fig.16: The tunneling rate $\gamma$ (blue solid curve) and the total probability $P(t)$ (black dashed curve) from the ground state at $g\rightarrow\infty$.
\end{center}

\subsection{4.3. Spectrum rearrangement}
In this Subsection we analyze the spectrum rearrangement which occurs for our anharmonic trap $V^{\textsf{sw}}$ at  the transition of the special point $g=0$ of non-interacting atoms with increasing $g$ from small negative to small positive values or vice versa. In Fig.5 we labeled the upper and lower energy branches at the negative side of the coupling constant $g$ with quantum numbers ($2,0$) and ($0,2$) correspondingly. These quantum numbers conserve only for the negative side of the coupling $g$. When these branches cross the points $g=0$ the nodal structures of these states rearrange and the states - ($2,0$) and ($0,2$) - interchange between each other.

With decreasing the anharmonic parameter $\alpha$ (see Fig.17) the effect becomes more pronounced.
\begin{center}
\includegraphics[scale=0.25]{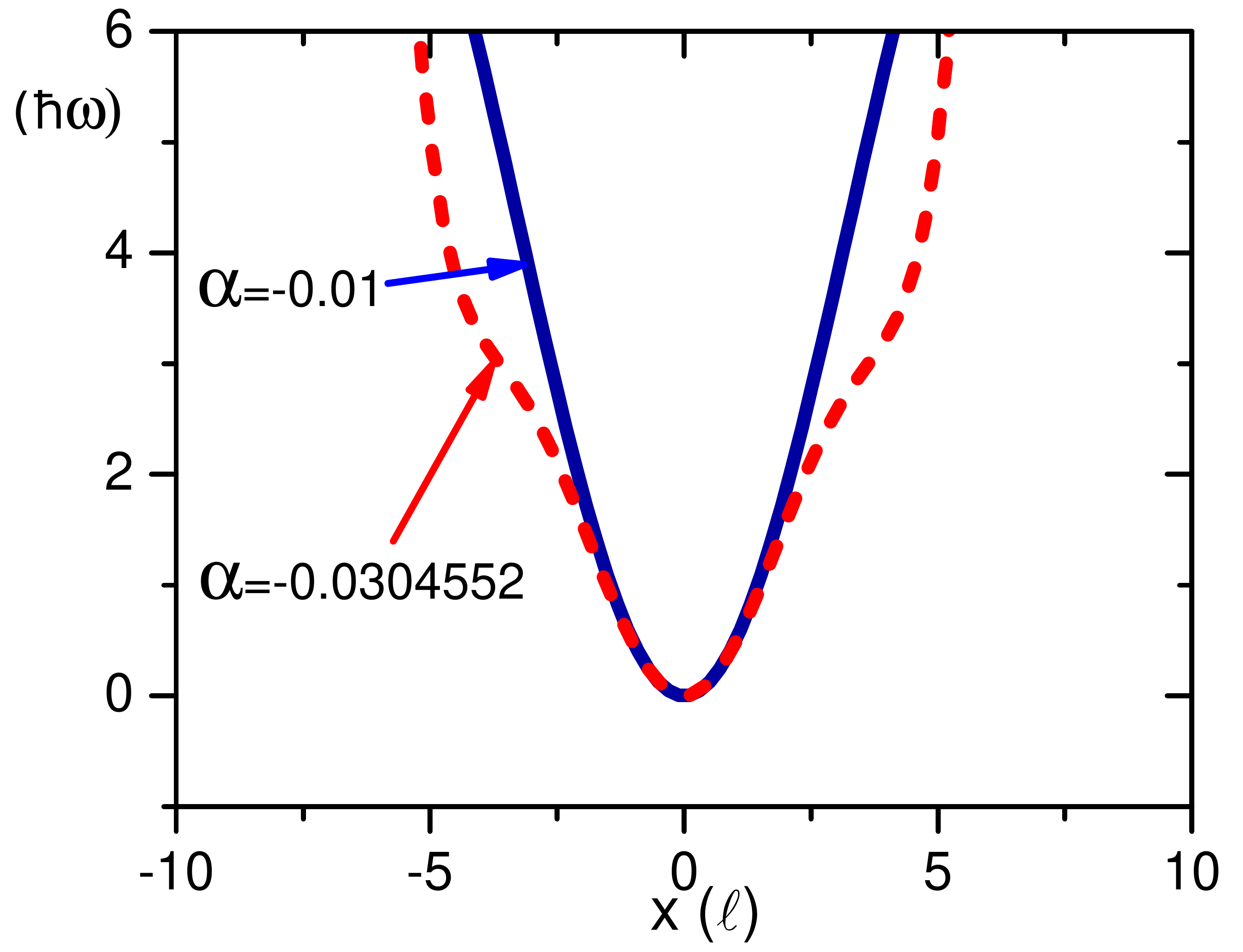}\\
Fig.17: The confining potential $V^{(6)}(x)$ for two values of the anharmonic parameter $\alpha$: $\alpha=-0.0304552$ (red dashed line) and $\alpha=-0.01$ (blue solid line).
\end{center}

Fig.18 shows the calculated energy levels of the pair ($2,0$) and ($0,2$) of the first excited states for $\alpha=-0.01$.
\begin{center}
\includegraphics[scale=0.25]{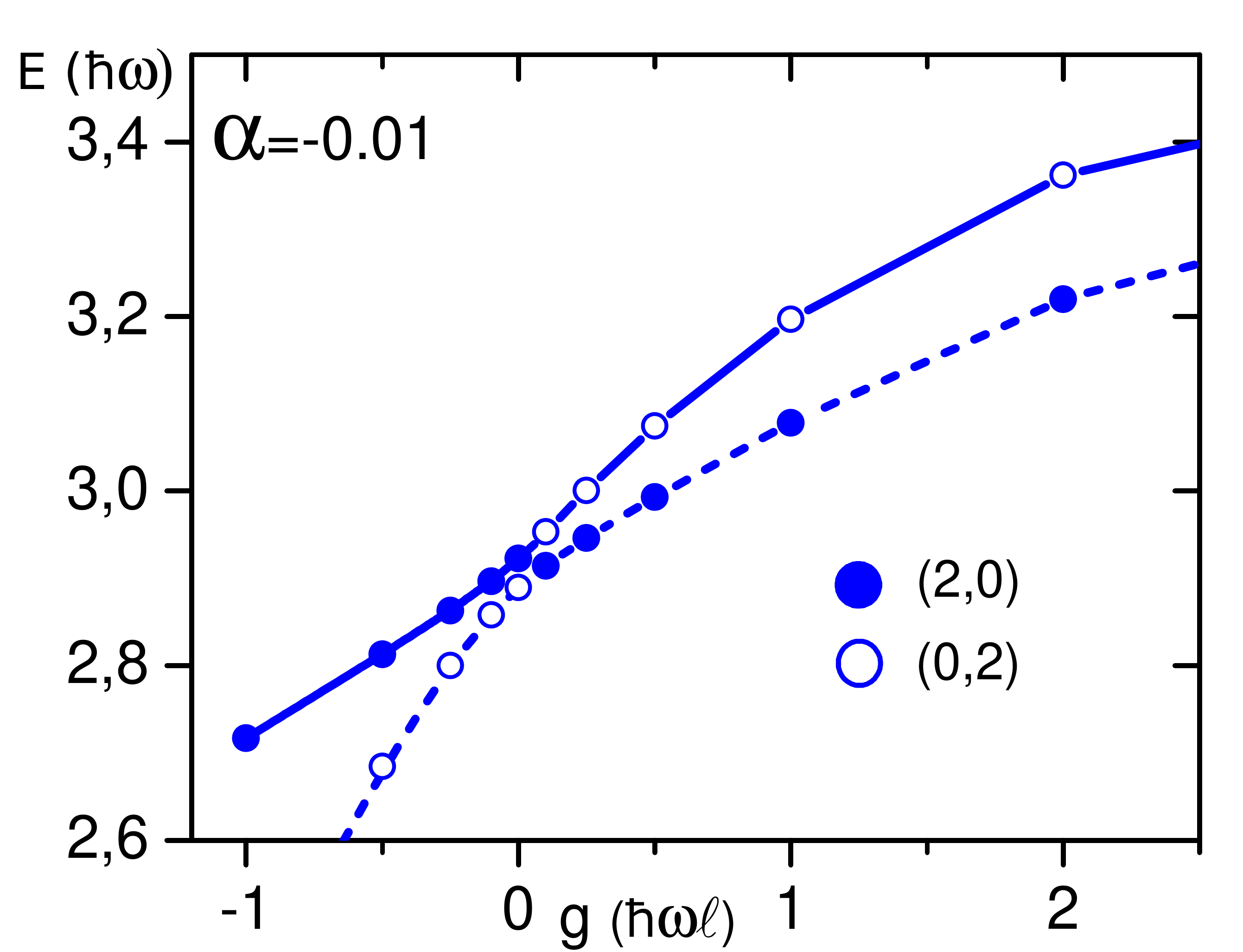}\\
Fig.18: Energy levels of the pair (2,0) (closed circles) and (0,2) (open circles) of the first excited states for the confining potential $V^{(6)}(x_j)$ with $\alpha=-0.01$.
\end{center}
The corresponding wave functions for the two cases - $\alpha=-0.01$ and $\alpha=-0.0304552$ of the upper and lower branches, given in Fig.18 and Fig.5 correspondingly, are presented in Fig.19 and Fig.20, which are plotted as functions of relative $x=x_1-x_2$ and center-of-mass $y=(x_1+x_2)/2$ variables.

From Figs.19 and 20 we see how the wave functions change their nodal structures when they cross the point $g=0$: the nodal structure of the upper branch's eigenfunction interchange between ($2,0$) and ($0,2$) states; similar effect occurs for the nodal structure of lower branch's eigenfunction, only vice versa.
The similar effect of the spectrum rearrangement for two atoms confined in 3D anharmonic trap was observed in calculations of \cite{sala}

We have also observed some kind of rearrangement of the nodal structure of the calculated wave function when crossing the point $g/(\hbar\omega\ell)\simeq3$ at $\alpha=-0.0304552$ (see Figs.19 and 20). However, due to strong interatomic coupling $g$ and considerable anharmonic parameter $\alpha$ we have strong mixing of the states with different quantum numbers here and cannot interpret the effect as a simple transition from one pure quantum state to another one like near the point $g=0$.

The calculated dependence of the initial atomic distribution (the probability density $|\psi(x,y,t=0)|^2$) on the coupling constant $g$ clarifies the monotonic increase of the tunneling rate from the lower energy branch of the excited states with the increase of $g$: with the increase of $g$ the maximums of the probability density moves closer to the regions (3) in Fig.9 (see Fig.20). The dependence on $g$ for the tunneling rate from the upper energy branch of the excited states has non-monotonic character due to more complicated dependence of the probability density $|\psi(x,y,t=0)|^2$ on $g$: with deviation to the left or right from the point $g=0$ the maximums of the probability density first approach to the regions (3) in Fig.9, but then start to move from the regions (3) to (4) and finally stabilize (see Fig.19).

\onecolumngrid
\hspace{-1cm}
\begin{tabular}{cccc}
\multicolumn{4}{c}{$\alpha=-0.01$}\\
g=-1 & g=1 & g=5 & g=20\\
\includegraphics[width=.27\textwidth]{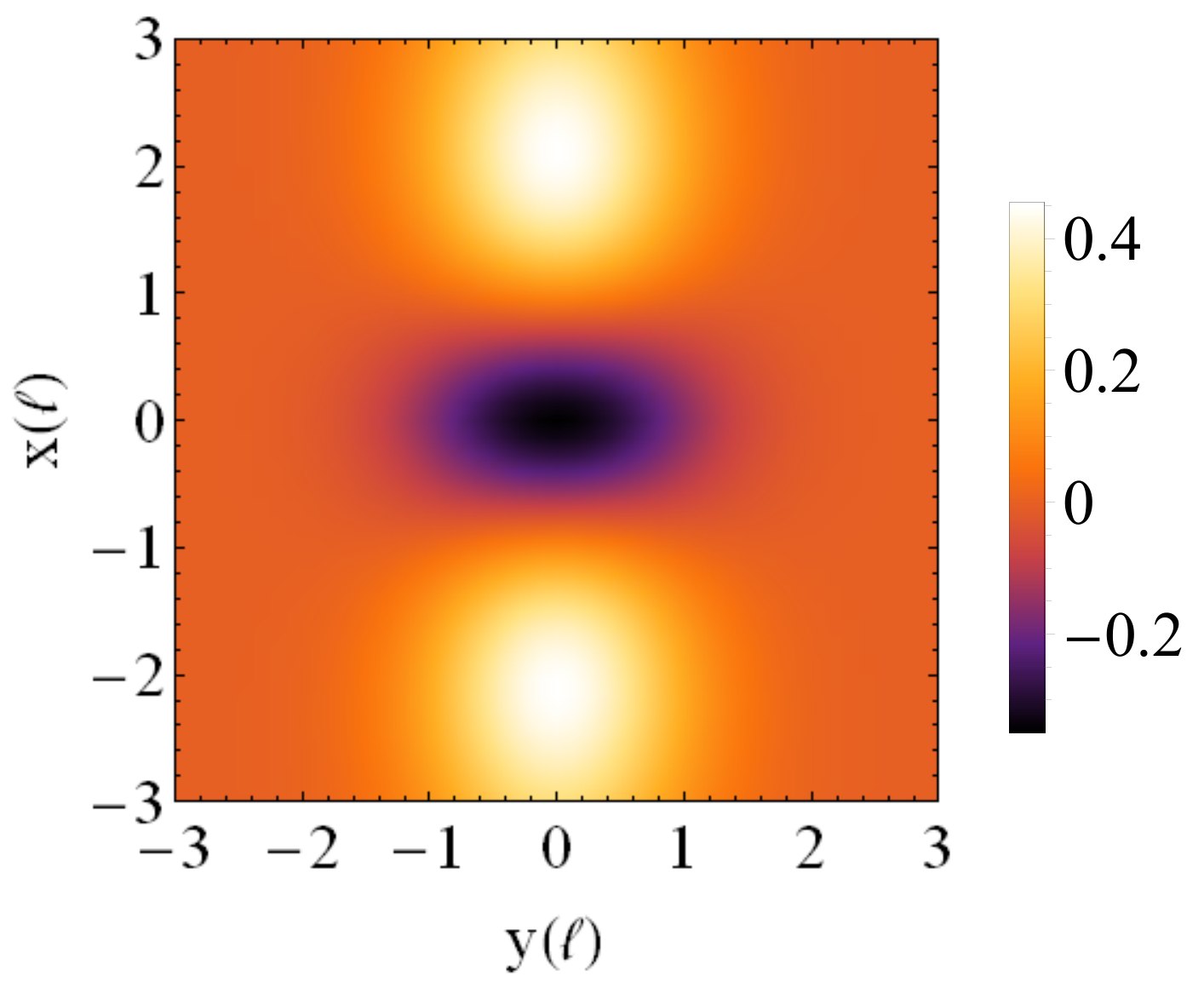} &
\includegraphics[width=.27\textwidth]{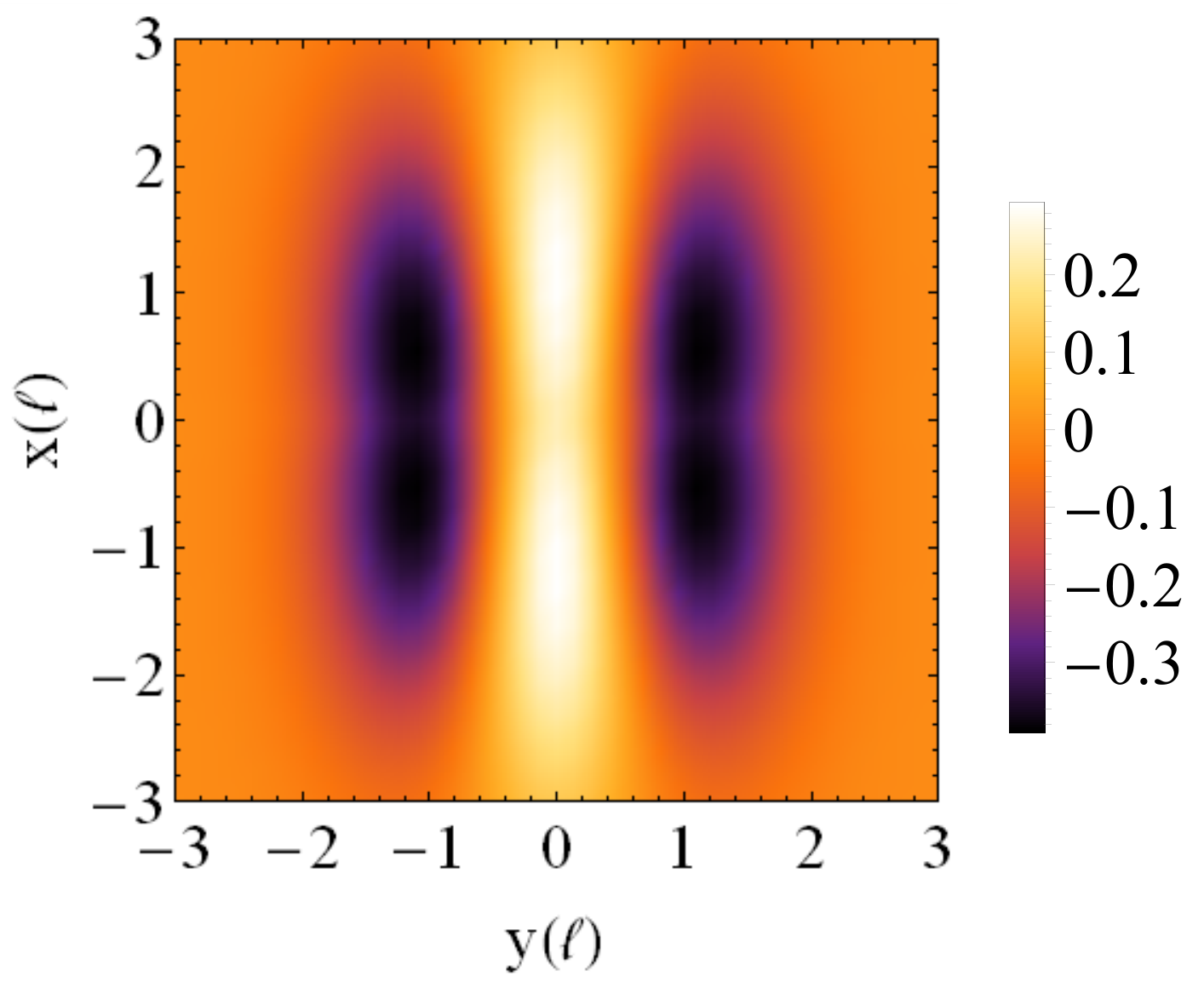} &
\includegraphics[width=.27\textwidth]{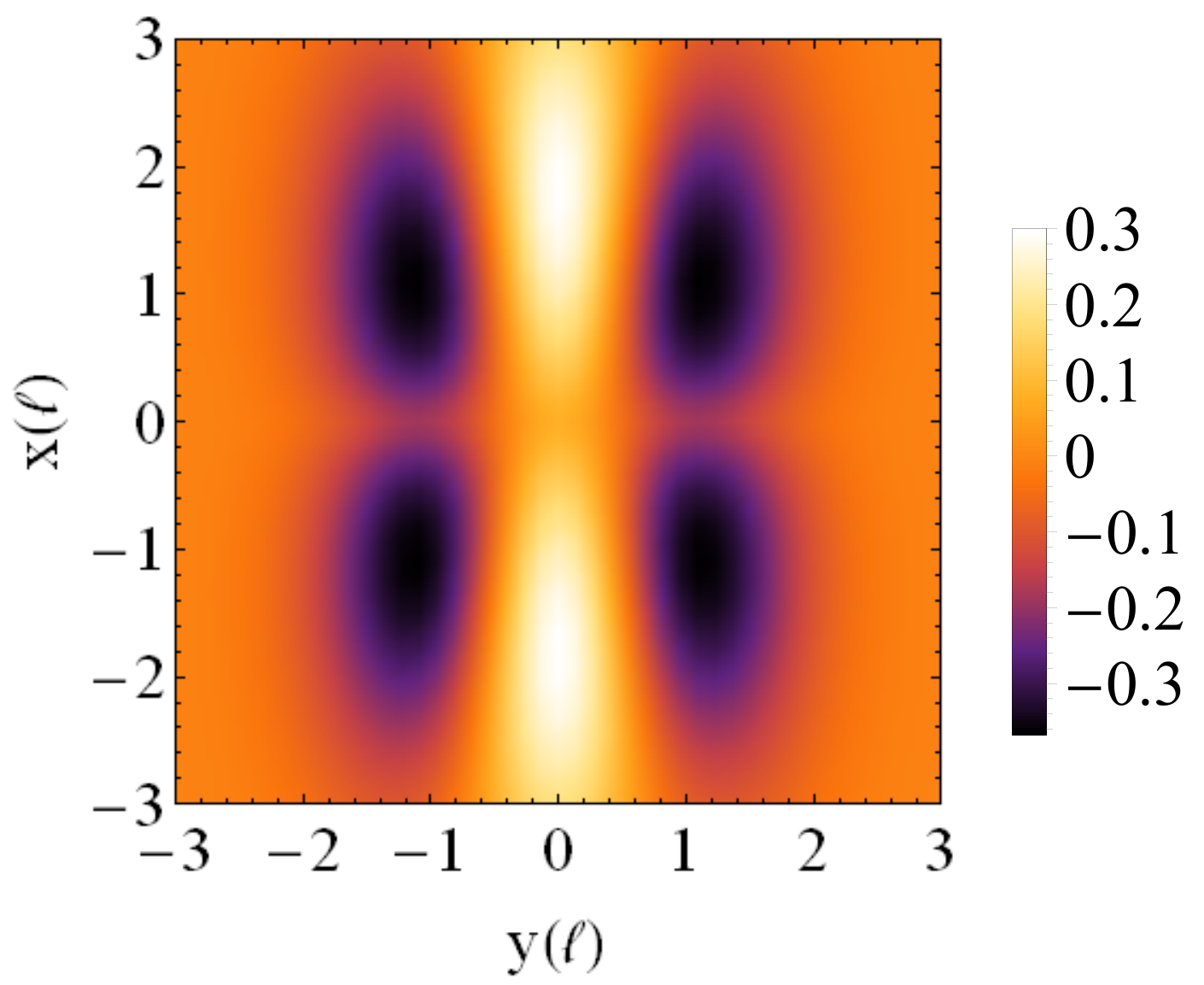} &
\includegraphics[width=.27\textwidth]{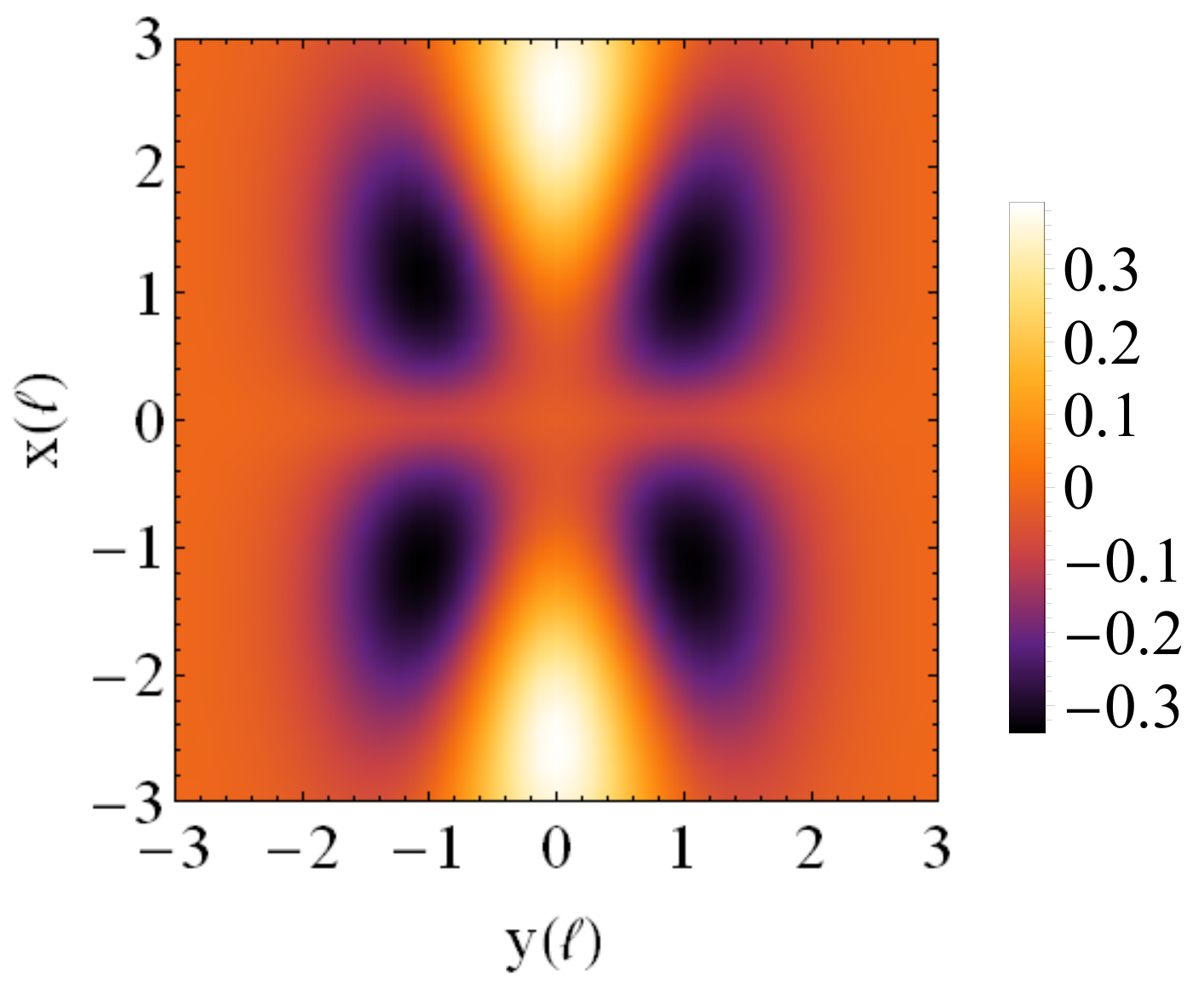} \\
\multicolumn{4}{c}{$\alpha=-0.0304552$}\\
g=-1 & g=1 & g=5 & g=20\\
\includegraphics[width=.27\textwidth]{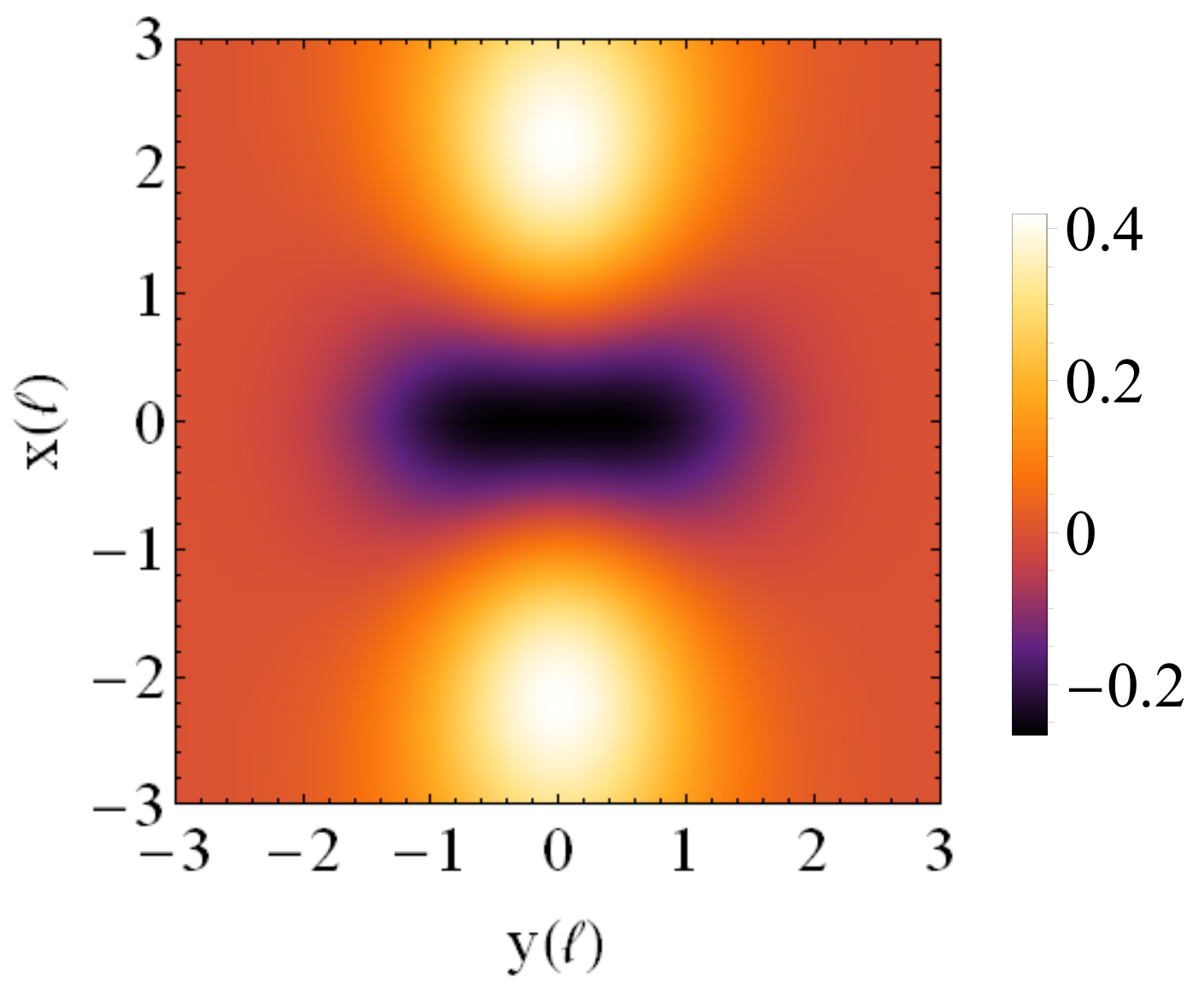} &
\includegraphics[width=.27\textwidth]{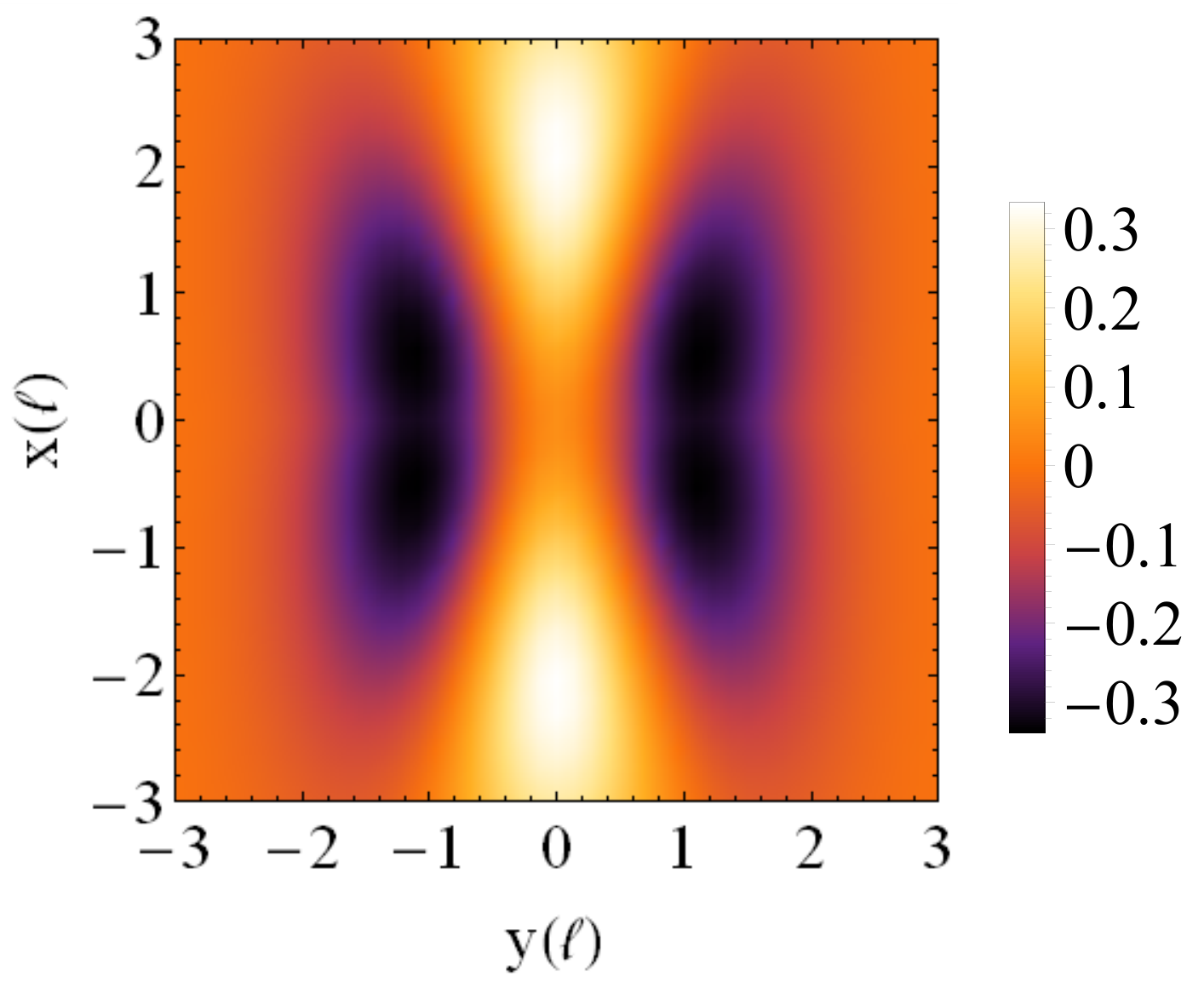} &
\includegraphics[width=.27\textwidth]{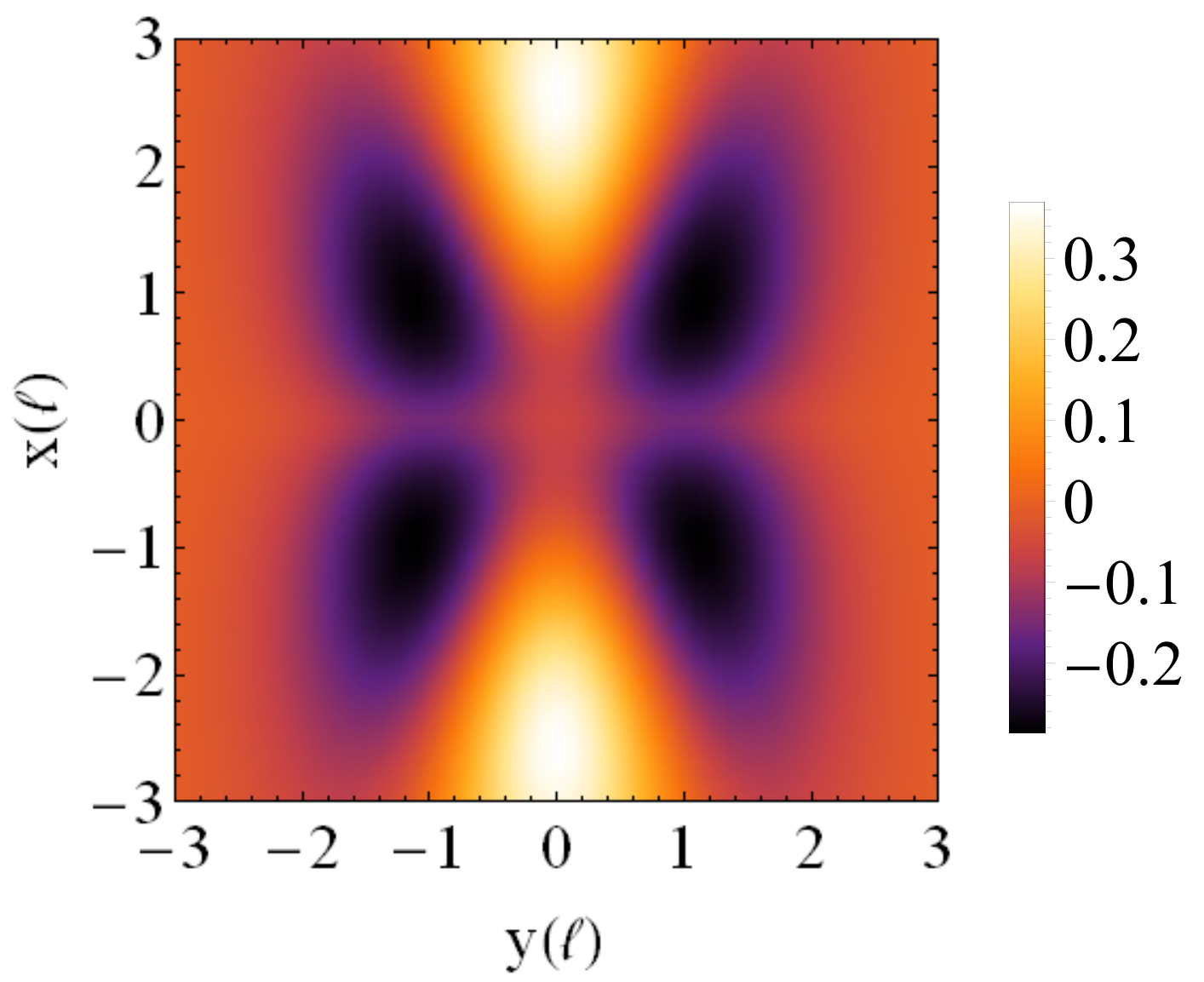} &
\includegraphics[width=.27\textwidth]{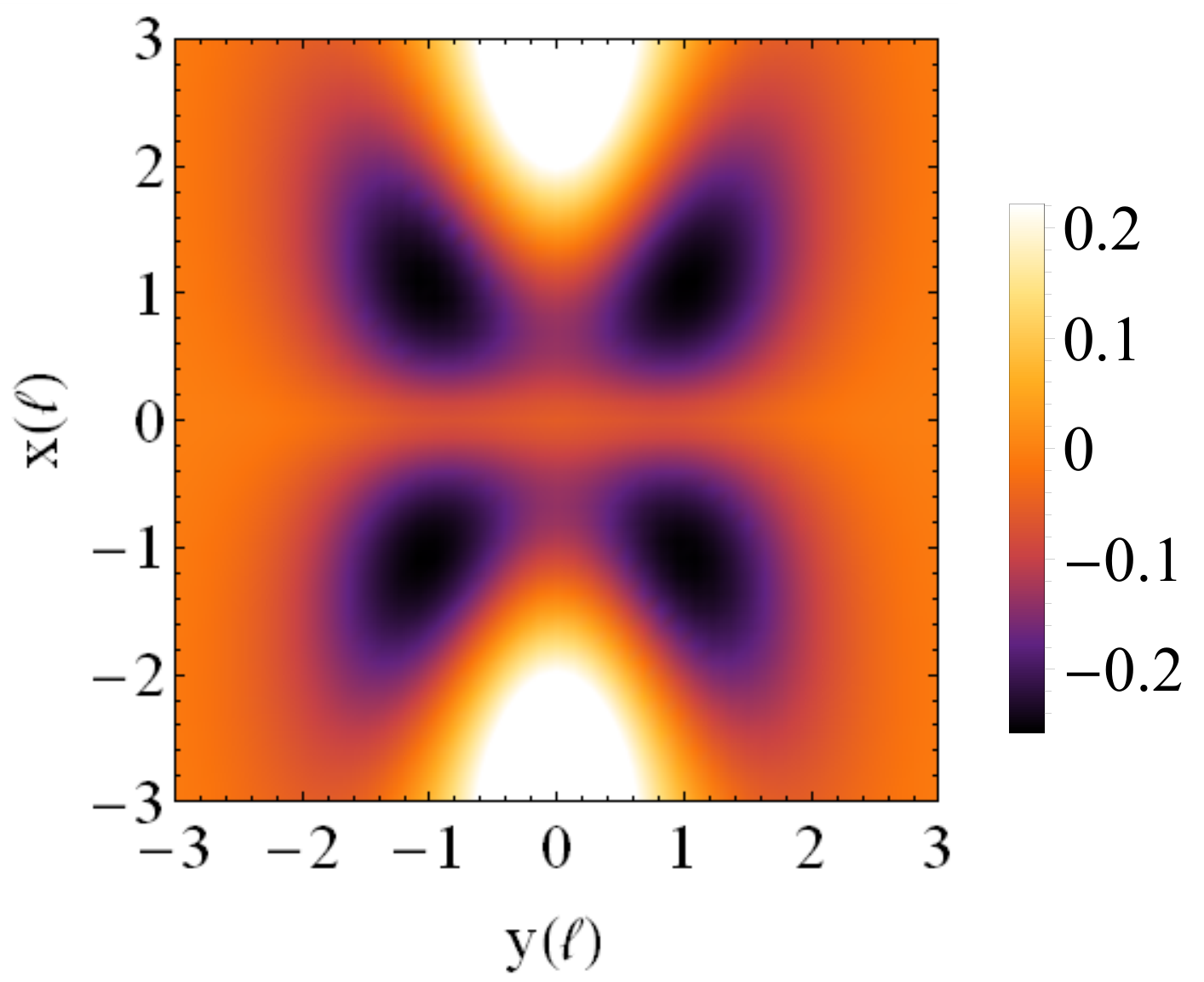} \\
\end{tabular}
\begin{center}
Fig.19: Wave function, $\psi(x,y,t=0)$ (in $\ell^{-1}$ units), of the upper energy branch in Fig.17. Coupling strength $g$ is in $\hbar\omega\ell$ units.
\twocolumngrid
\onecolumngrid
\end{center}
\hspace{-1cm}
\begin{tabular}{cccc}
\multicolumn{4}{c}{$\alpha=-0.01$}\\
g=-1 & g=1 & g=5 & g=20\\
\includegraphics[width=.27\textwidth]{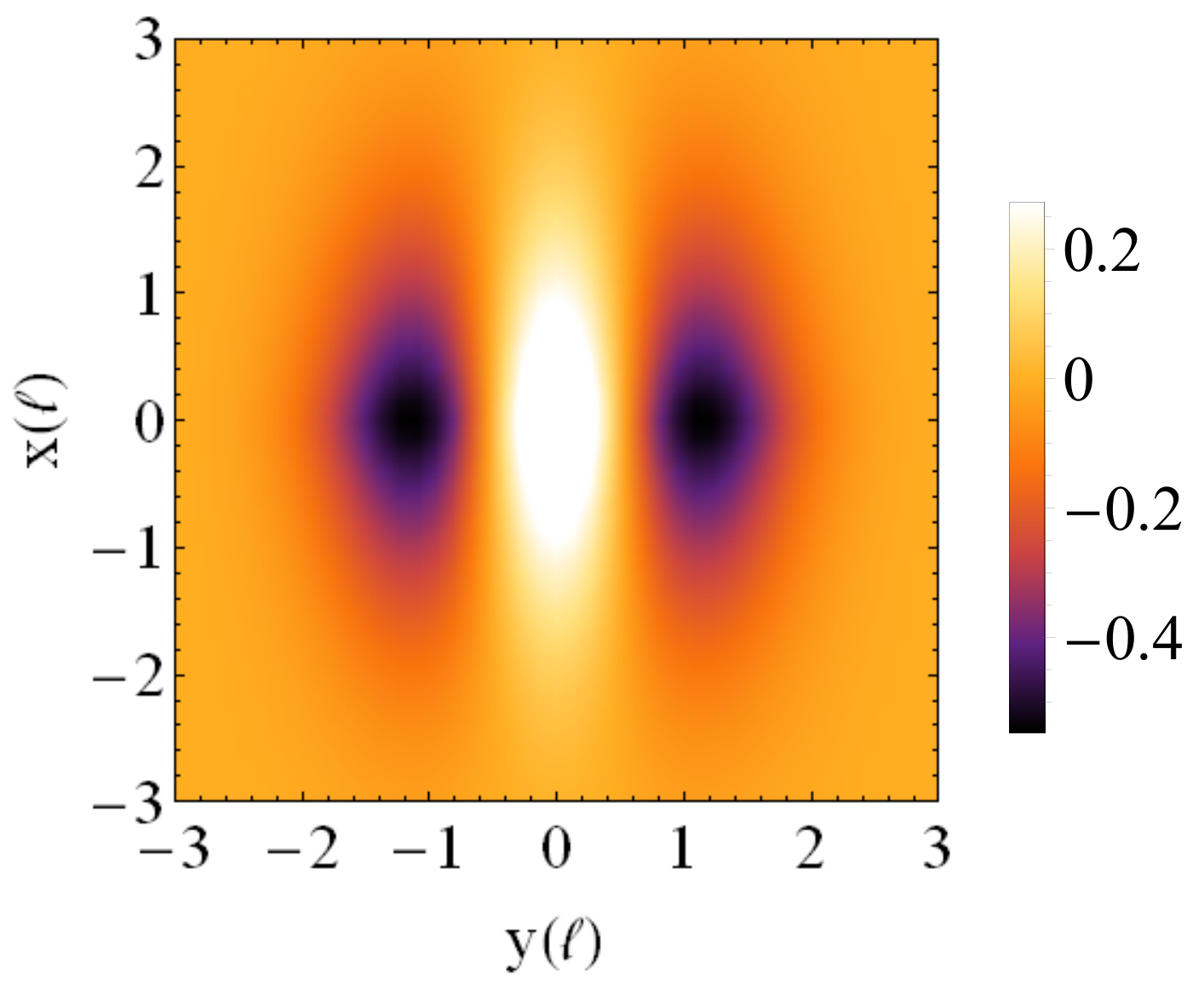} &
\includegraphics[width=.27\textwidth]{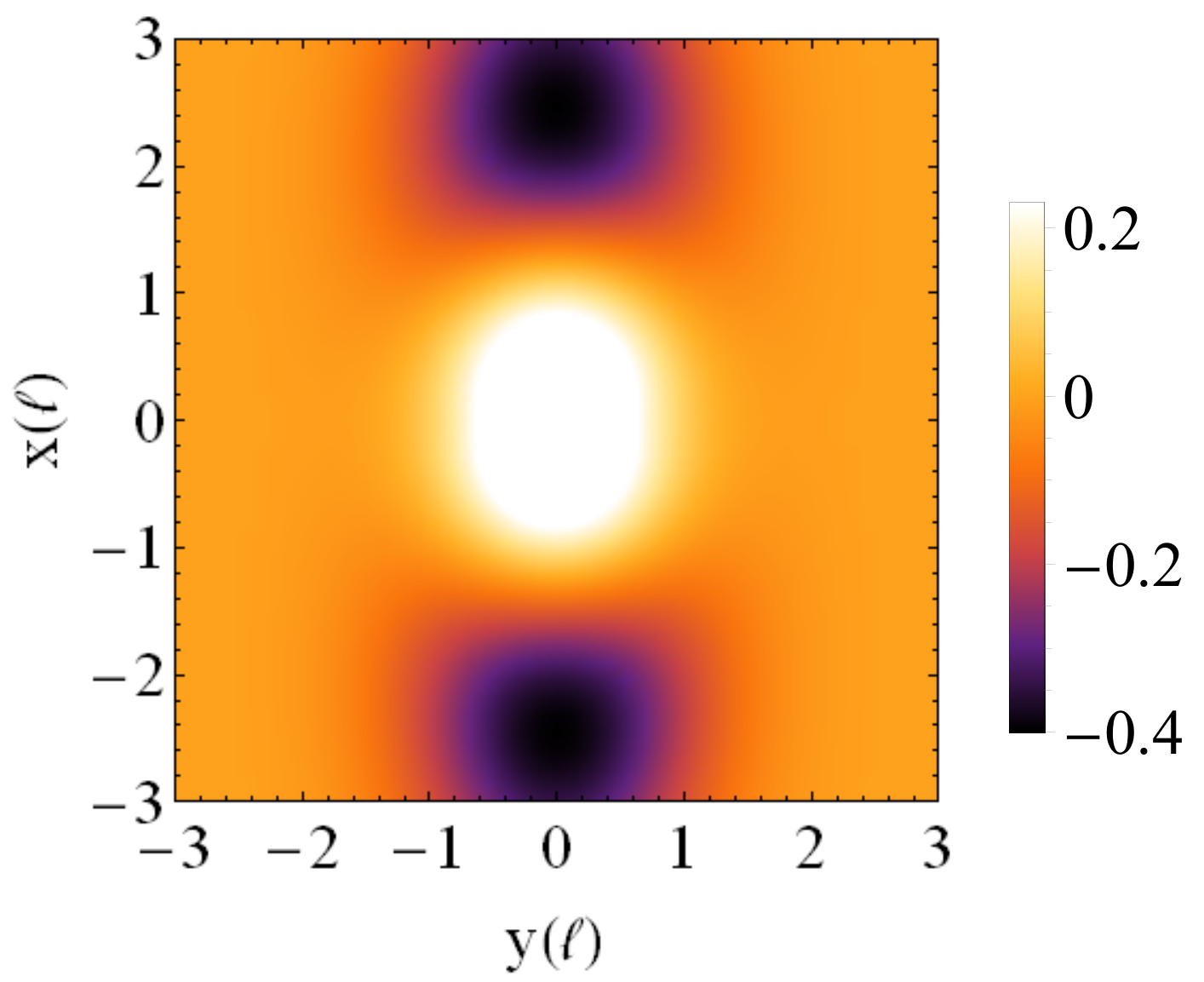} &
\includegraphics[width=.27\textwidth]{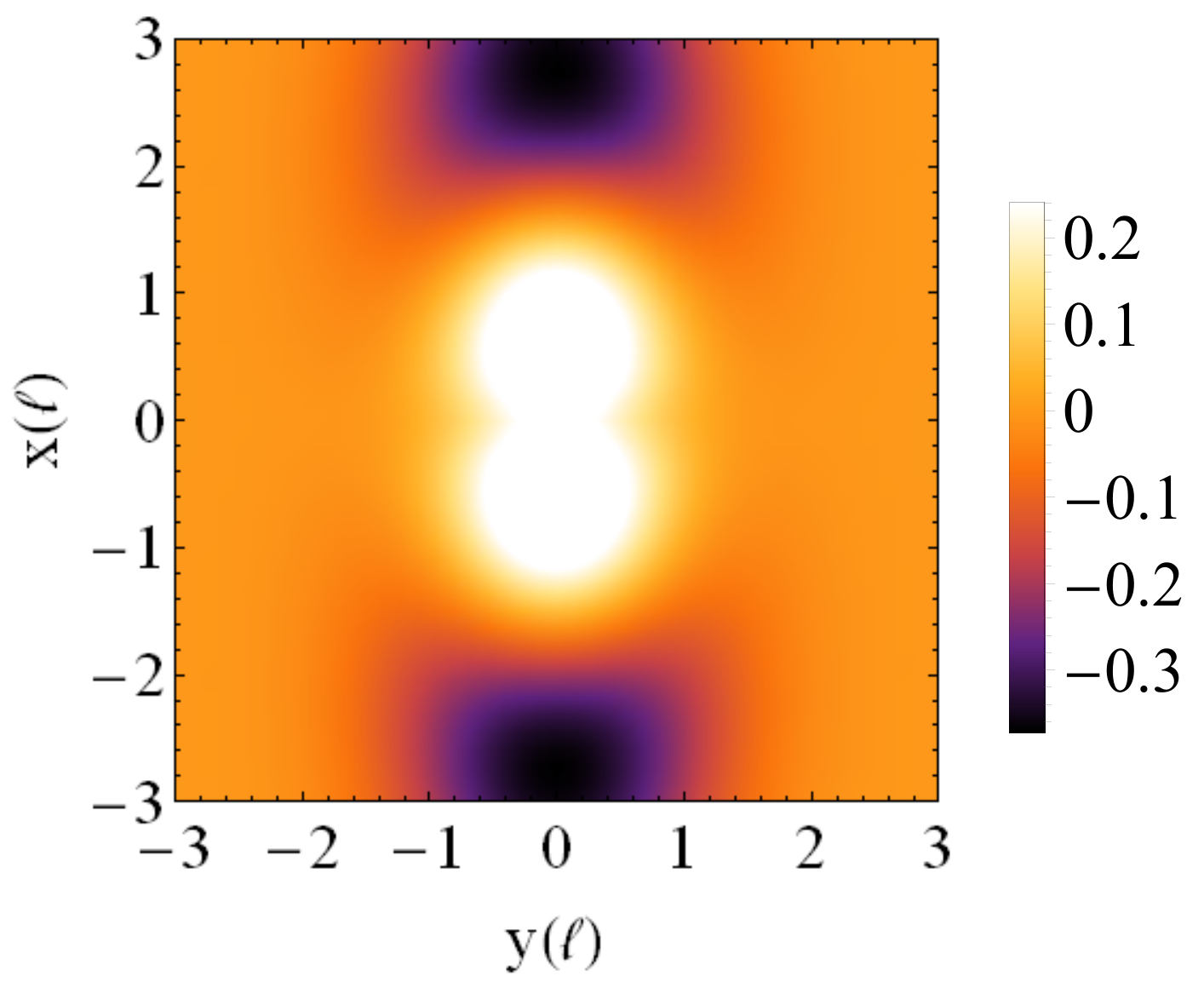} &
\includegraphics[width=.27\textwidth]{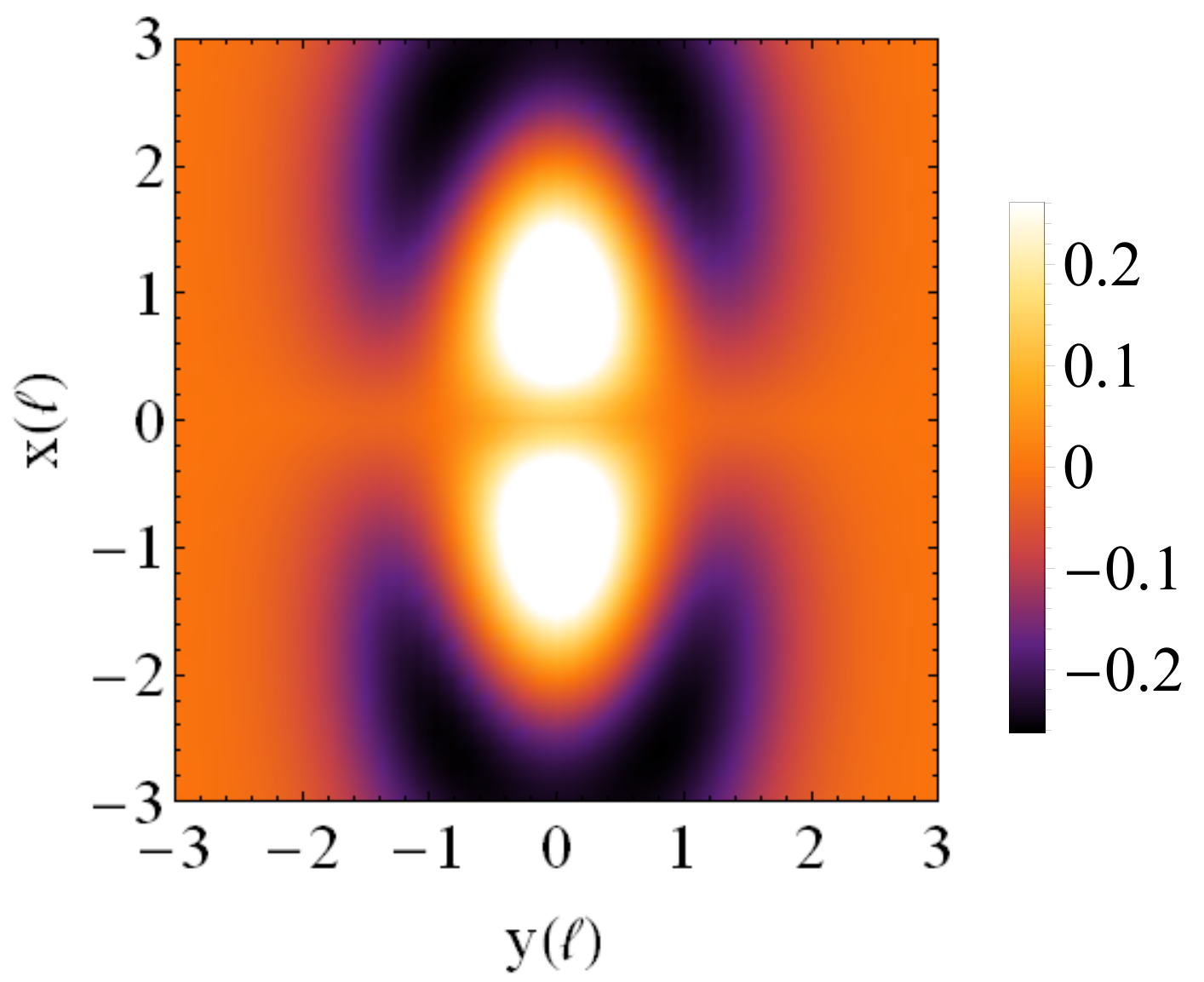} \\
\multicolumn{4}{c}{$\alpha=-0.0304552$}\\
g=-1 & g=1 & g=5 & g=20 \\
\includegraphics[width=.27\textwidth]{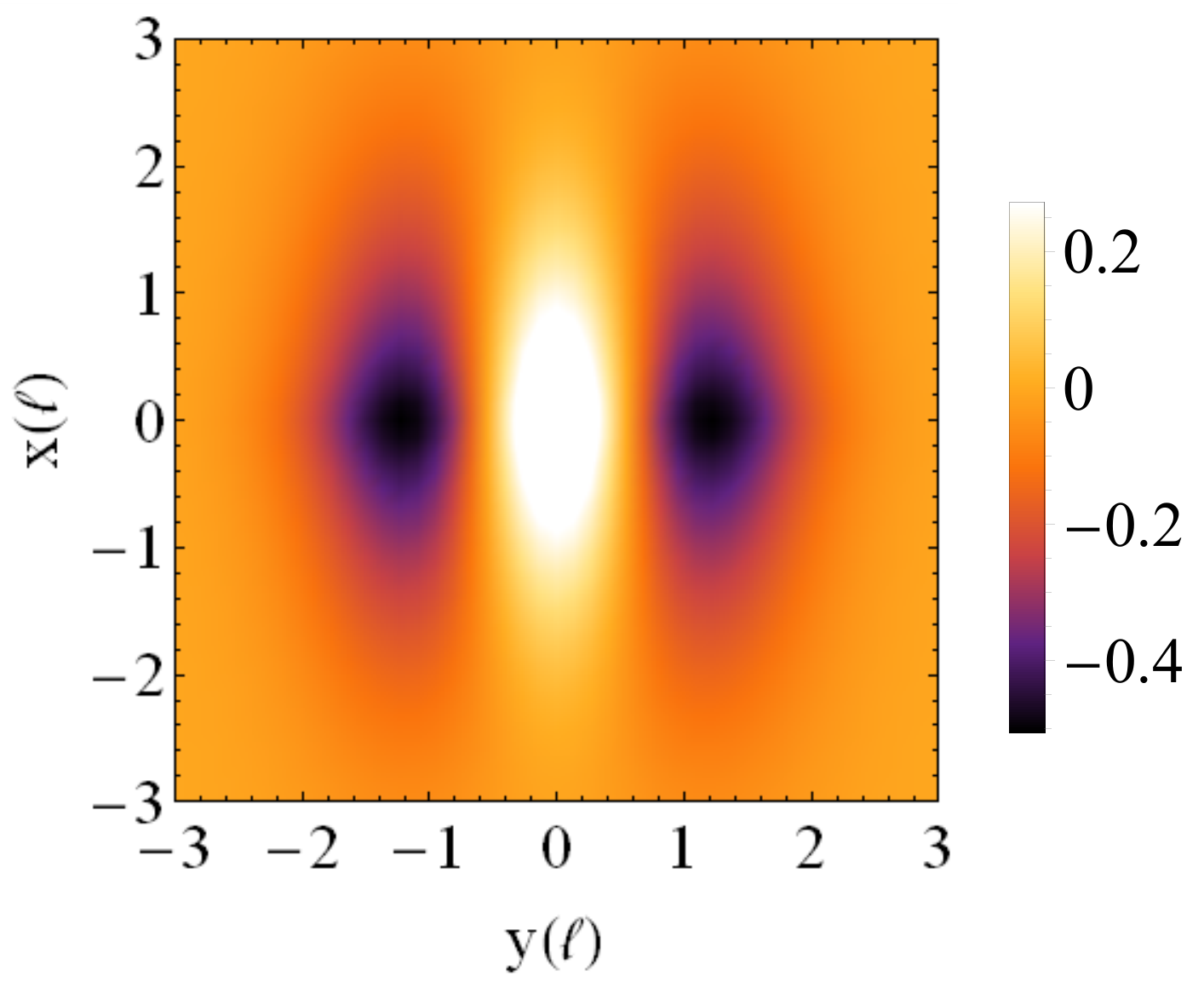} &
\includegraphics[width=.27\textwidth]{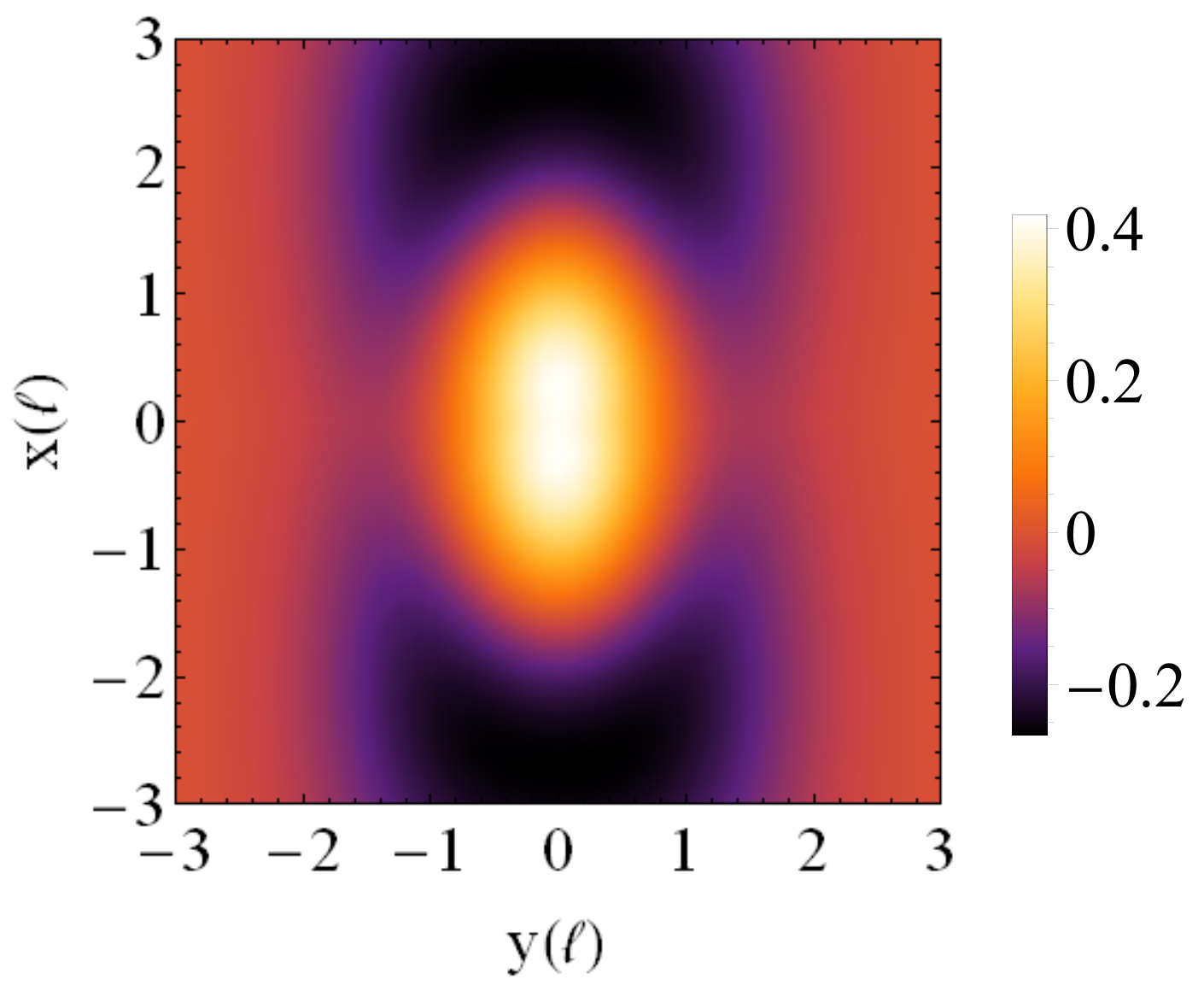} &
\includegraphics[width=.27\textwidth]{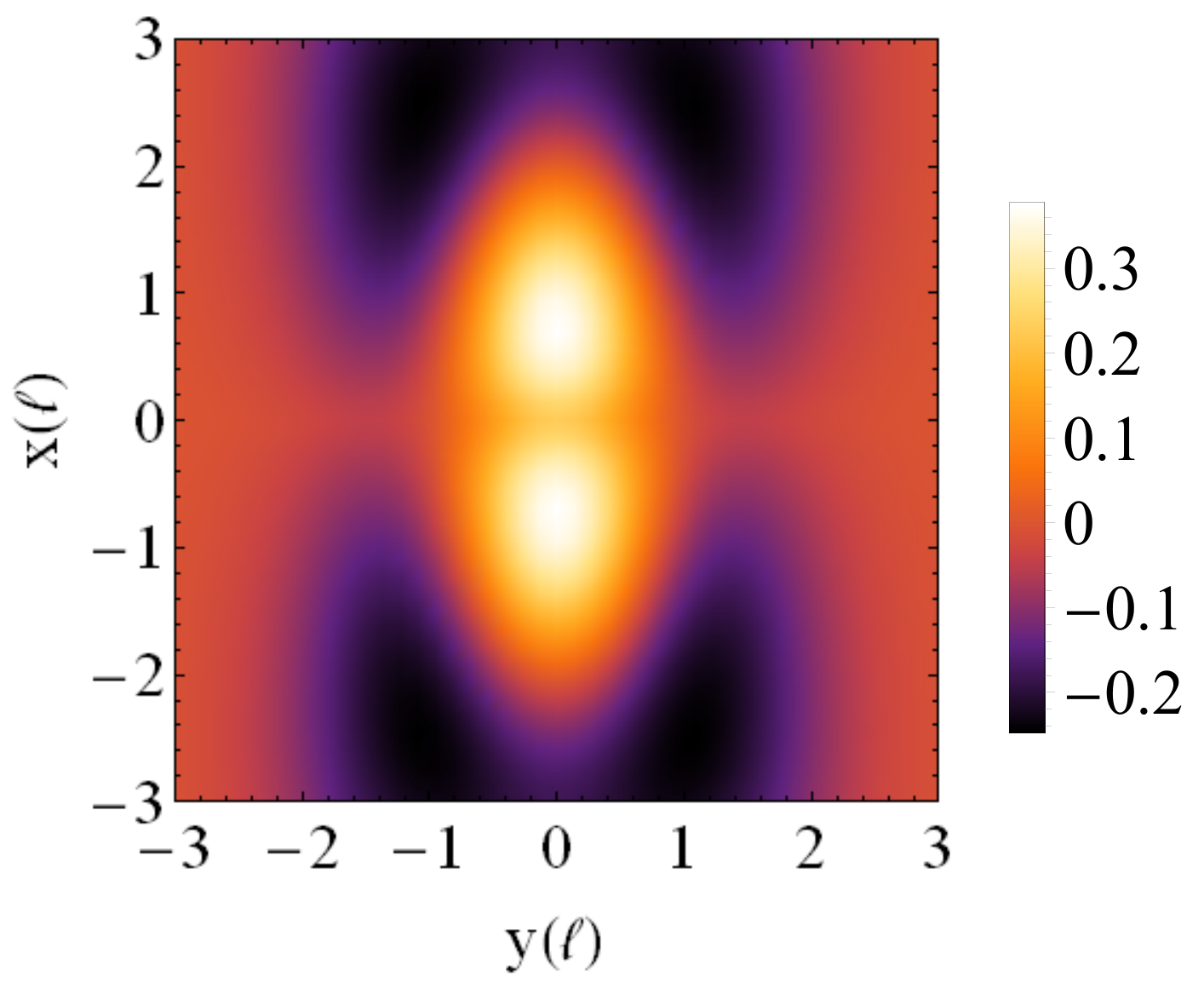} &
\includegraphics[width=.27\textwidth]{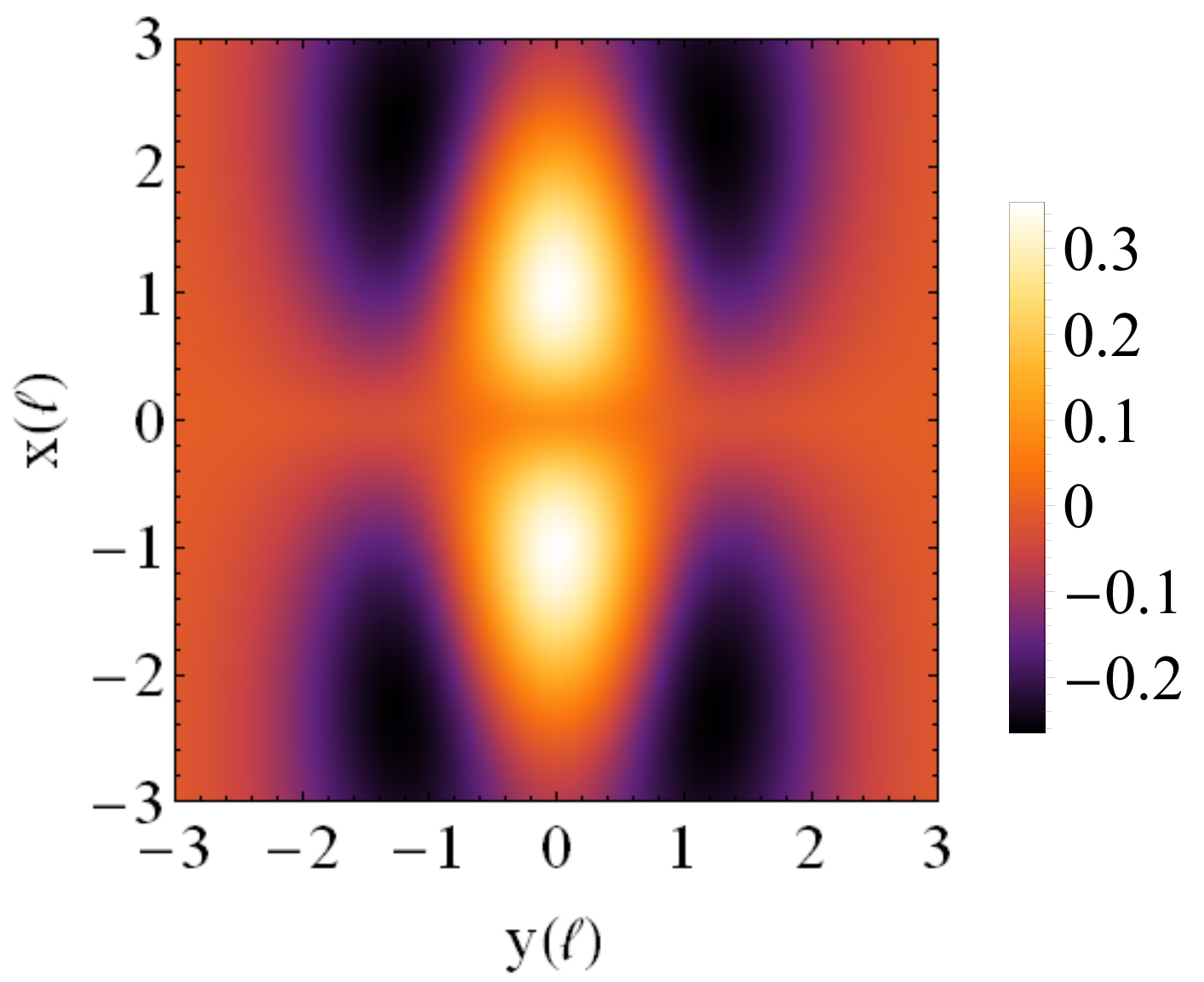} \\
\end{tabular}
\begin{center}
Fig.20: Wave function, $\psi(x,y,t=0)$ (in $\ell^{-1}$ units), of the lower energy branch in Fig.17. Coupling strength $g$ is in $\hbar\omega\ell$ units.
\end{center}
\twocolumngrid

\section{V. Conclusion}
We have investigated the tunneling dynamics of a 1D system of two interacting atoms confined in the anharmonic trap (2.5). We have calculated the tunneling rates $\gamma$ from the three lowest atomic bound states as a function of the coupling constant $g$ for different values $\alpha$ of the trap anharmonicity. It was found that in the tunneling from the upper energy branch of the excited states, $\gamma$ behaves non-monotonically and the sequential particle channel dominates in the tunneling. Note, that the domination of the sequential tunneling was also observed in the box-shaped potential model from the ground state of a rectangular potential well\cite{hunn}. When the atoms are initially in the lower energy branch of the excited states, $\gamma$ grows very fast with increasing of the coupling strength $g$ at the beginning of the tunneling. Then, it was found, that the tunneling passes in new regime(s) with more slow tunneling rate(s) due to the competition of the tunneling with the transition to the upper energy branch of the excited states and to the ground state. When the atoms tunnel from the ground state, $\gamma$ grows slowly and monotonically with increasing $g$.

We have also analyzed a rearrangement $(0,2)\rightleftarrows(2,0)$ of the spectrum in the limit $g\rightarrow \pm 0$ of noninteracting atoms with the exchange of the wave function nodal structure. More complicated rearrangement of the nodal structure of the calculated wave function of the confined pair of atoms was found when crossing the point $g/(\hbar\omega\ell)\simeq3$.

The developed computational scheme can be extended to technically more complicated, but close to current and planning experiments\cite{zuern,zuern2}, a problem about the tunneling from quasi-1D cigar-like and quasi-2D pancake-like traps. Including the spin dynamics into the model for tunneling process is another actual problem which can be investigated with the developed technique.

The authors are very grateful to P. Schmelcher, V. Pupyshev, Yu.V. Popov, and  S.I. Vinitsky for very helpful comments and fruitful discussions.

The work was supported by the Grant of the Plenipotentiary Representative of the Republic of Kazakhstan to JINR.


\begin{thebibliography}{99}

\bibitem{zuern} G. Z\"urn, A. N. Wenz, S. Murmann, A. Bergschneider, T. Lompe, and S. Jochim, \textit{Pairing in Few-Fermion Systems with Attractive Interactions}, Phys. Rev. Lett. \textbf{111}, 175302 (2013).

\bibitem{beinke} R. Beinke, S. Klaiman, L. S. Cederbaum, A. I. Streltsov, and O. E. Alon , \textit{Many-Body Tunneling Dynamics of Bose-Einstein Condensates and Vortex States in Two Spatial Dimensions}, Phys. Rev. A \textbf{92}, 043627 (2015).

\bibitem{lode} A. U. J. Lode, A. I. Streltsov, O. E. Alon, H.-D. Meyer, and L. S. Cederbaum, \textit{Exact Decay and Tunnelling Dynamics of
Interacting Few-Boson Systems}, J. Phys. B \textbf{42}, 044018 (2009).

\bibitem{lode1} A. U. J. lode, \textit{Tunneling Dynamics in Open Ultracold Bosonic Systems}, Springer Theses (Springer Cham Heidelberg New York Dordrecht London) 2015.

\bibitem{nesterenko} V. O. Nesterenko, A. N. Novikov, and E. Suraud, \textit{Transport of the Repulsive Bose-Einstein Condensate in a Double-Well Trap: Interaction Impact and Relation to the Josephson Effect}, Laser Phys. \textbf{24}, 125501 (2014).

\bibitem{zuern2} G. Z\"{u}rn, F. Serwane, T. Lompe, A. N. Wenz, M. G. Ries, J. E. Bohn, and S. Jochim, \textit{Fermionization of Two Distinguishable Fermions}, Phys. Rev. Lett. \textbf{108}, 075303 (2012).

\bibitem{rontani} M. Rontani \textit{Tunneling Theory of Two Interacting Atoms in a
Trap}, Phys. Rev. Lett. \textbf{108}, 115302 (2012).

\bibitem{rontani2} M. Rontani \textit{Pair tunneling of two atoms out of a trap}, Phys. Rev. A \textbf{88}, 043633 (2013).

\bibitem{gharashi} S.E. Gharashi and D. Blume \textit{Tunneling dynamics of two interacting one-dimensional particles}, Phys. Rev. A \textbf{92}, 033629 (2015).

\bibitem{melezhik} V.S. Melezhik, \textit{A computational method for quantum dynamics of a three-dimensional atom in strong fields}, WE-Heraeus-Seminar (Germany) ``Atoms and Molecules in Strong External Fields'', (Plenum, New-York and London, 1998) pp.89-94.

\bibitem{krass} P. M. Krassovitskiy, F. M. Pen'kov, \textit{Izvestiya Akademii Nauk. Ser. Fizicheskaya} \textbf{79}(7):1041-1046, (2015)

\bibitem{gusev} A. A. Gusev, S. I. Vinitsky, O. Chuluunbaatar, V. L. Derbov, A. G\'{o}\'{z}d\'{z} and P. M. Krassovitskiy, \textit{Metastable states of a composite system tunneling through repulsive barriers}, Theor Math Phys 186: 21, (2016)

\bibitem{maruyama} T. Maruyama, T. Oishi, K. Hagino, and H. Sagawa, \textit{Time-dependent approach to many-particle tunneling in one dimension}, Phys. Rev. C \textbf{86}, 044301 (2012)

\bibitem{scamps} G. Scamps and K. Hagino, \textit{Multidimensional fission model with a complex absorbing potential}, Phys. Rev. C \textbf{91}, 044606 (2015).

\bibitem{hunn} S. Hunn, K. Zimmermann, M. Hiller, and A. Buchleitner, \textit{Tunneling Decay of Two Interacting Bosons in an Asymmetric Double-Well Potential: A Spectral Approach}, Phys. Rev. A \textbf{87}, 043626 (2013).

\bibitem{peng} S.-G. Peng, H. Hu, X.-J. Liu and P.D. Drummond, \textit{Confinement-Induced Resonances in Anharmonic Waveguides}, Phys. Rev. A \textbf{84}, 043619 (2011).

\bibitem{bloch} I. Bloch, J. Dalibard, and W. Zwerger, \textit{Many-body physics with ultracold gases}, Rev. Mod. Phys. \textbf{80}, 885, (2008).

\bibitem{ishmukh} I.S. Ishmukhamedov, D.T. Aznabayev, and S.A. Zhaugasheva, \textit{Two-body atomic system in a one-dimensional anharmonic trap: The energy spectrum}, Phys. Part. Nuclei Lett. \textbf{12}: 680 (2015).

\bibitem{haller} E. Haller, M. J. Mark, R. Hart, J. G. Danzl, L. Reichsollner,
V. Melezhik, P. Schmelcher, and H.-C. Nagerl,
\textit{Confinement-induced resonances in low-dimensional quantum systems}, Physical Review
Letters, \textbf{104}, 153203, (2010).

\bibitem{marchuk} G.I. Marchuk, \textit{Methods of Numerical Mathematics}, Springer-Verlag, New York (1975), Sec. 4.3.3

\bibitem{melezhik2} V.S. Melezhik, J.I. Kim and P. Schmelcher, \textit{Wave Packet Dynamical Analysis of Ultracold Scattering in Cylindrical Waveguides}, Phys. Rev. A \textbf{76}, 053611 (2007).

\bibitem{melezhik3} V.S. Melezhik, \textit{Mathematical Modeling of Ultracold Few-Body Processes in Atomic Traps}, EPJ Web Conf. \textbf{108}, 01008 (2016).

\bibitem{meyer} U. V. Riss and H.-D. Meyer, \textit{Calculation of resonance energies and widths using the complex absorbing potential method}, J. Phys. B \textbf{26}, 4503 (1993).

\bibitem{kulander} J. L. Krause, K. J. Schafer, and K. C. Kulander, \textit{Calculation of photoemission from atoms subject to intense laser fields}, Phys. Rev. A \textbf{45}, 4998 (1992).

\bibitem{olshanii} M. Olshanii, \textit{Atomic Scattering in the Presence of an External Confinement and a Gas of Impenetrable Bosons}, Phys. Rev. Lett., \textbf{81}, 938 (1998).

\bibitem{busch} T. Busch, B. Englert, K. Rzazewski, and M. Wilkens, \textit{Two Cold Atoms in a Harmonic Trap}, Found. Phys. \textbf{28}, 549 (1998).

\bibitem{ishmukh2} I.S. Ishmukhamedov, D.S. Valiolda, and S.A. Zhaugasheva, \textit{Description of ultracold atoms in a one-dimensional geometry of a harmonic trap with a realistic interaction}, Phys. Part. Nuclei Lett. \textbf{11}: 238 (2014).


\bibitem{grish} S. Grishkevich and A. Saenz, \textit{Theoretical description of two ultracold atoms in a single site of a three-dimensional optical lattice using realistic interatomic interaction potentials}, Phys. Rev A \textbf{80}, 013403 (2009).

\bibitem{sala} S. Sala and A. Saenz, \textit{Theory of inelastic confinement-induced resonances due to the coupling of center-of-mass and relative motion}, Phys. Rev. A \textbf{94}, 022713 (2016).



\end{thebibliography}
\end{document}